\documentclass[usenatbib]{mnras}

\usepackage[english]{babel}
\usepackage[utf8x]{inputenc}
\usepackage[T1]{fontenc}
\usepackage{pdflscape}
\usepackage[normalem]{ulem}

\DeclareGraphicsExtensions{  .pdf, .png}

\usepackage{amsmath}
\usepackage{graphicx}
\usepackage[colorinlistoftodos]{todonotes}

\newcommand{\remove}[1]{}

\newcommand{\add}[1]{#1}

\title[
HELP catalogue of prime extragalactic fields
]{HELP: A catalogue of 170 million objects, selected at 0.36---4.5 $\mu$m, from 1270 deg.$^{2}$ of prime extragalactic  fields}
\author[R. Shirley et al.]{
\parbox{\linewidth}{
Raphael Shirley,$^{1,2}$\thanks{E-mail: rshirley@iac.es }
Yannick Roehlly,$^{1,3,4}$
Peter D Hurley,$^{1}$
Veronique Buat,$^{3}$
María del Carmen Campos Varillas,$^{1}$
Steven Duivenvoorden,$^{1}$
Kenneth J Duncan,$^{5}$
Andreas Efstathiou,$^6$
Duncan Farrah,$^{7, 8}$
Eduardo González Solares,$^{9}$
Katarzyna Małek,$^{3, 10}$
Lucia Marchetti,$^{11, 12, 13}$
Ian McCheyne,$^{1}$
Andreas Papadopoulos,$^6$
Estelle Pons,$^{9}$
Roberto Scipioni,$^1$
Mattia Vaccari,$^{12, 13}$
and Seb Oliver.$^{1}$ 
}
\\
\\
$^{1}$Astronomy Centre, Department of Physics and Astronomy, University of Sussex, Brighton, UK, BN1 9QH\\
$^{2}$Instituto de Astrofísica de Canarias, E-38205 La Laguna, Tenerife, Spain;\\ Universidad de La Laguna, Dpto. Astrofísica, E-38206 La Laguna, Tenerife, Spain\\
$^{3}$Aix Marseille Univ, CNRS, CNES, LAM, Marseille, France\\
$^{4}$Université de Lyon, ENS de Lyon, CNRS, Centre de Recherche en Astrophysique de Lyon UMR5574, 69230 Saint-Genis-Laval, France\\
$^{5}$Sterrewacht Leiden, Universiteit Leiden, Leiden, Netherlands\\
$^{6}$School of Sciences, European University Cyprus, Diogenes street, Engomi, 1516 Nicosia, Cyprus\\
$^{7}$Department of Physics and Astronomy, University of Hawaii, 2505 Correa Road, Honolulu, HI 96822, USA\\ 
$^{8}$Institute for Astronomy, 2680 Woodlawn Drive, University of Hawaii, Honolulu, HI 96822, USA\\
$^{9}$Institute of Astronomy, University of Cambridge, Madingley Road, Cambridge, UK\\
$^{10}$National Centre for Nuclear Research, ul. Ho$\dot{z}$a 69, 00-681 Warszawa, Poland\\
$^{11}$Department of Astronomy, University of Cape Town, 7701 Rondebosch, Cape Town, South Africa\\
$^{12}$Department of Physics and Astronomy, University of the Western Cape, 7535 Bellville, Cape Town, South Africa\\
$^{13}$INAF - Istituto di Radioastronomia, via Gobetti 101, 40129 Bologna, Italy\\
}

\pubyear{2019}

\begin{document}
\maketitle

\begin{abstract}
We present an optical to near-infrared selected astronomical catalogue covering 1270 deg.$^2$. This is the first attempt to systematically combine data from 23 of the premier extragalactic survey fields -- the product of a vast investment of telescope time. The fields  are those imaged by the \textit{Herschel} Space Observatory which form the \textit{Herschel} Extragalactic Legacy Project (HELP). Our catalogue of 170 million objects is constructed by a  positional cross match of 51 public surveys. This high resolution optical, near-infrared, and mid-infrared catalogue is designed for photometric redshift estimation, extraction of fluxes in lower resolution far-infrared maps, and spectral energy distribution modelling. It collates, standardises, and provides value added derived quantities including corrected aperture magnitudes and astrometry correction over the \emph{Herschel} extragalactic wide fields for the first time. $grizy$ fluxes are available on all fields with $g$ band data reaching $5\sigma$ point-source depths in a 2 arcsec aperture of 23.5, 24.4, and 24.6 (AB) mag at the 25th, 50th, and 75th percentiles, by area covered, across all HELP fields. It has $K$ or $K_s$ coverage over 1146 deg.$^2$ with depth percentiles of 20.2, 20.4, and 21.0 mag respectively. The IRAC Ch 1 band is available over 273 deg.$^2$ with depth percentiles of 17.7, 21.4, and 22.2 mag respectively. This paper defines the \emph{``masterlist''} objects for the first data release (DR1) of HELP. This large sample of standardised total and corrected aperture fluxes, uniform quality flags, and completeness measures provides large well understood statistical samples over the full \emph{Herschel} extragalactic sky.

\end{abstract}
 
 
\clearpage

\section{Introduction}
\label{sec:intro}
Galaxy catalogues play a fundamental role in modern astronomy and cosmology. Combining catalogues and images from different instruments is a significant challenge and will be increasingly important as deeper and wider surveys are conducted in the coming years. Huge efforts have gone in to creating large homogeneous data sets such as for the SXDS \citep{Furusawa:2008}, the Optical-NIR catalogue of the AKARI-NEP Deep Field \citep{Oi:2014}, COSMOS \citep[e.g.][]{Scoville:2007, Ilbert:2013, Laigle:2016}, and GAMA \citep{Driver:2011}. However, creating a consistent data set over the \emph{Herschel} extragalactic fields is a challenge due to the wide variety of projects that have studied these fields. Each survey is at a different depth, with different source extraction pipelines, different astrometric solutions and units used, and different quality in terms of seeing, point spread function consistencies and exposure times. In this paper we collate and combine a large number of public astronomical catalogues to produce a single catalogue of general use to the astronomical community.

Access to such data sets over a wide area extends the scientific value for e.g. the discovery of rare objects, reducing sampling variance, studying environmental factors and statistical studies in fine sub-samples of populations. Dedicated large-area surveys such as the Dark Energy Survey \citep[DES;][]{Abbott:2018} and the upcoming Large Synoptic Survey Telescope \citep[LSST;][]{Ivezic:2008} typically only provide five or six optical bands. To access a wide multi-wavelength range for physical modelling and to exploit the deepest data requires the combination of many data sets from different telescopes.  The premier extra-galactic fields\footnote{AKARI-NEP, AKARI-SEP, Bootes, CDFS-SWIRE, COSMOS, EGS, ELAIS-N1, ELAIS-N2, ELAIS-S1, GAMA-09, GAMA-12, GAMA-15, HDF-N, Herschel-Stripe-82, Lockman-SWIRE, NGP, SA13, SGP, SPIRE-NEP, SSDF, xFLS, XMM-13hr, and XMM-LSS.} represent many hundreds of nights of the best ground based telescopes and thousands of hours of space telescope time and yet have never been effectively collected together.

The \textit{Herschel}\footnote{\textit{Herschel} Space Observatory \citep{Pilbratt:2010}} Extragalactic Legacy Project (HELP\footnote{\url{https://herschel.sussex.ac.uk/} }) (Oliver et al., in preparation) brings together key multi-wavelength surveys. It homogenises them;  exploits prior information from the optical, near-infrared and mid-infrared surveys to deblend the low-resolution long wavelength maps using {\sc xid+} \citep{Hurley:2016}; provides well-calibrated photometric redshifts \citep{Duncan:2018, Duncan:2017}; and conducts energy-balanced spectral energy distribution (SED) modelling through {\sc cigale} \citep{Buat:2018,Malek:2018} to provide physical parameter estimations. 

The first stage in this process is to compile all the ancillary data that is available across the HELP fields into a \emph{``masterlist''} of objects. The principal aim is  to produce a catalogue with consistent photometry from the best photometric catalogues that are available.  A secondary aim is to characterise the depth of the selected data to enable meaningful statistical analysis with standard empirical measures such as luminosity functions.

Here, we describe in detail the production and the extensive validation, flagging and characterisation of the \emph{masterlist} across the full HELP area. We do this in part by discussing an example field, the European Large Area ISO Survey field North 1 \citep[ELAIS-N1;][]{Oliver:2000}, in detail. Although we use an example field for discussing the combination methods we provide detailed quantitative measures of the depths and number counts for all the fields individually and combined over the entire area.

The \emph{masterlist} presented here is the basis for the first public HELP data release (HELP DR1) and defines the HELP identifiers and positions which will be used in subsequent data products for DR1. 

Full details of decisions, validation and characterisation are available in the {\sc Jupyter} notebooks\footnote{\url{https://github.com/H-E-L-P/dmu_products} } \citep{Kluyver:2016} that were used to run all the {\sc Python} code used in the pipeline and are made available as part of the HELP documentation. 

This work is of general use for galaxy formation studies by providing larger statistical samples than have been available before. The provision of tools to model selection functions in particular permit the combination of narrow deep fields with wide shallow fields to investigate both the bright and faint end of the luminosity function simultaneously. It facilitates the use of large numbers of different surveys for such research. For its specific purpose of deblending far-infrared maps, computing photometric redshifts, and modelling spectral energy distributions of galaxies, it is of critical importance. Constructing a prior list of objects for the application of Bayesian forced photometry with {\sc xid+} requires a uniformly defined selection which is well correlated with far-infrared flux. This catalogue can be consistently used for such a purpose across all \emph{Herschel} imaging, from the heavily observed and deep COSMOS field to the hundreds of square degrees of HATLAS-SGP.

The format of this paper is as follows. First, in section~\ref{sec:input_data} the HELP fields are described and an overview of the various input data sets is given and a detailed overview of our example, pilot field, ELAIS-N1, is given to demonstrate the method of data combination. The data reduction pipeline is described in section~\ref{sec:pipeline} with some examples from the example field, ELAIS-N1. Validation of the data and the depth maps used to quantify survey depths are described in sections~\ref{sec:validation} and \ref{sec:depth_maps} respectively. In section~\ref{sec:depth_overview} we compare the depths and number accounts across all the fields both combined and individually, for critical detection bands. Finally, the data presented is summarised in section~\ref{sec:conclusions}.

\section{The input data sets}
\label{sec:input_data}

In this section, we first give an overview of the fields and their location on the sky. Then, using ELAIS-N1 as a demonstrative example, we describe the method of data combination that has been applied on all the fields. There are 51 public surveys combined in this work. A full list of all these surveys with information regarding the instruments and bands used is given in table~\ref{table:full-pristine}.

\subsection{HELP fields}
\begin{figure*}
\centering
\includegraphics[width=1.0\textwidth]{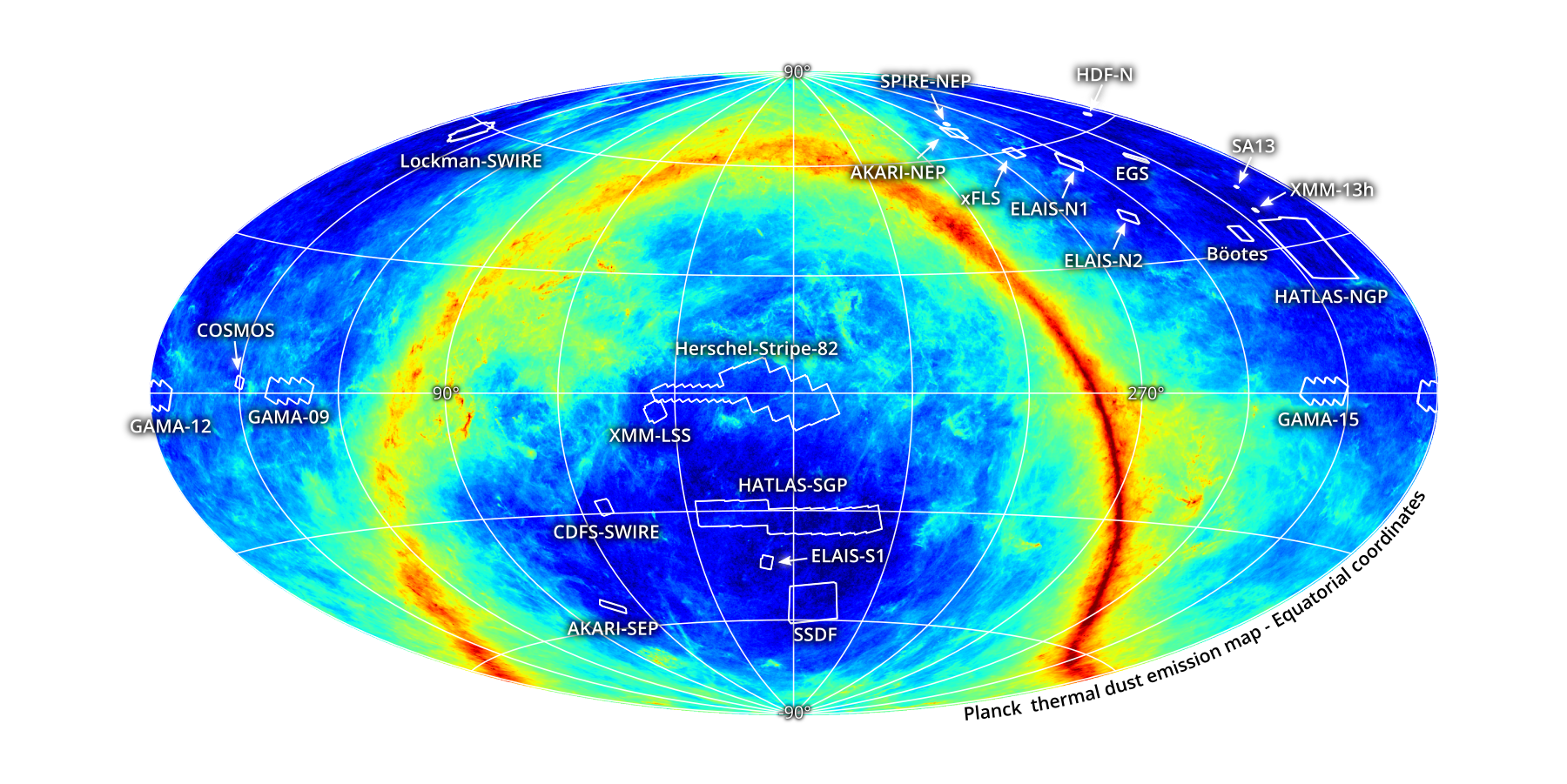}
\caption{\label{fig:fields} The boundaries of the HELP fields overlayed on the Planck Galactic thermal dust emission map. Field areas range from less than one deg.$^2$ (HDF-N, SA13, SPIRE-NEP, and XMM-13hr) to several hundred deg.$^2$ (e.g. HATLAS-SGP).
}
\end{figure*}

There are 23 HELP fields. Figure~\ref{fig:fields} shows their location on the sky. Creating statistical samples from these various fields in order to study galaxy formation and evolution first requires collecting data across the fields. In order for these samples to be scientifically useful they must have a well described selection function. It is therefore necessary to combine and manipulate data in a reproducible manner and to quantify depth on every patch of sky.

The data available on a given field is highly variable in terms of number of public surveys, the bands measured by those surveys, the coverage of each band on the fields, and the depths of those surveys. On XMM-LSS for example there are 23 surveys available although their respective coverages are not uniform across the field. On HATLAS-SGP however, covering 295 deg.$^2$, there are just 6 surveys and no mid-infrared coverage by \textit{Spitzer}. This is where the depth maps presented in section~\ref{sec:depth_maps} become crucial for understanding selection effects leading to differences in effective area depending on a given sample selection. Also in section~\ref{sec:depth_maps}, we will discuss the depths available from specific filters on specific instruments in terms of the standard broad band filters: $u$, $g$, $r$, $i$, $z$, $y$, $J$, $H$, $K$, $K_s$, IRAC Ch 1 ($3.6\mu$m - `$i$1'), IRAC Ch 2 ($4.5\mu$m - `$i$2'), IRAC Ch 3 ($5.6\mu$m - `$i$3'), and IRAC Ch 4 ($8.0\mu$m - `$i$4'). In total there are 150 specific filters used in the HELP data. 29 of these are narrow band filters. In the following sections we will describe the typical broad band filters used. All the filters have transmission curves available in the database, as illustrated in figure~\ref{fig:filters}. The filters were taken from the Spanish Virtual Observatory (SVO) \citep{Rodrigo:2012,Rodrigo:2017} or from the original survey databases. These are corrected for atmospheric extinction and CCD quantum efficiency, and are as used in all subsequent HELP data processing.

\subsubsection{Optical data}
We define the optical region as between 0.36---1.0~$\mu$m. All of the HELP fields have some coverage by optical surveys in the $g$, $r$, $i$, $z$, and $y$ bands. Some areas of HELP are also covered by the broad band $u$ filter in addition to narrow bands. Later in the paper we will describe tools for determining what area is covered by each of these bands. General descriptions of all the optical data and its coverage of HELP area are given in appendix~\ref{sec:appendix}.

\subsubsection{Near-infrared data}
We define the near-infrared region as between 1.0---3.0~$\mu$m. 1146 deg.$^2$ of HELP is covered by either $K$, or $K_s$. All HELP fields have some coverage excluding the smaller fields (< 10 deg.$^2$) ELAIS-N2, SA13, SPIRE-NEP, xFLS, and XMM-13hr. There are fluxes from one or more of the $J$, $H$, $K$, and $Ks$ bands from 10 instruments. Full descriptions of the telescopes, instruments and bands are given in appendix~\ref{sec:appendix}.

\subsubsection{Mid-infrared data}
We define mid-infrared data as between 3.0---10.0~$\mu$m. Photometry from this part of the spectrum exclusively comes from the IRAC camera on the \textit{Spitzer} space telescope. We have measurements in IRAC bands over 273 deg.$^2$ of HELP.

A summary of the datasets available across all HELP fields is given in table~\ref{table:full-pristine}. They are each based on different selection criteria corresponding to features of the instrument, observations, and extraction software and have varying depths and coverage. 

\subsubsection{Multi-wavelength Catalogues}
On some HELP fields we have multi-wavelength catalogues produced with forced photometry from $K$ selected catalogues (where the positions from a $K$-band catalogue are used to extract fluxes from the other imaging bands), or from other methods such as stacked images as in the COSMOS2015 catalogue \citep{Laigle:2016} (where a $\chi^2$ sum of the four near-infrared bands and the optical $z$ is used to extract positions and forced photometry is then conducted on those positions). Where catalogues contain narrow bands in addition to the common broad bands, we include all bands available.

\subsection{An example field in detail: ELAIS-N1}
\label{sec:input_data_en1}

In this section we will demonstrate some of the general data properties and the combination method on an example field. We chose the ELAIS-N1 field to be representative of the HELP coverage in general. ELAIS-N1 was the first \textit{Spitzer} Wide InfraRed Extragalactic \citep[SWIRE;][]{Lonsdale:2003} field observed. This field is used to demonstrate the full HELP pipeline on areas with InfraRed Array Camera (IRAC) coverage, important for the HELP deblending with {\sc xid+} \citep{Hurley:2016}. It has previously been used to test another part of the HELP pipeline \citep{Malek:2018} and for early studies using HELP data \citep{Ocran:2017}. 

\begin{figure*}
\centering
\includegraphics[width=0.9\textwidth]{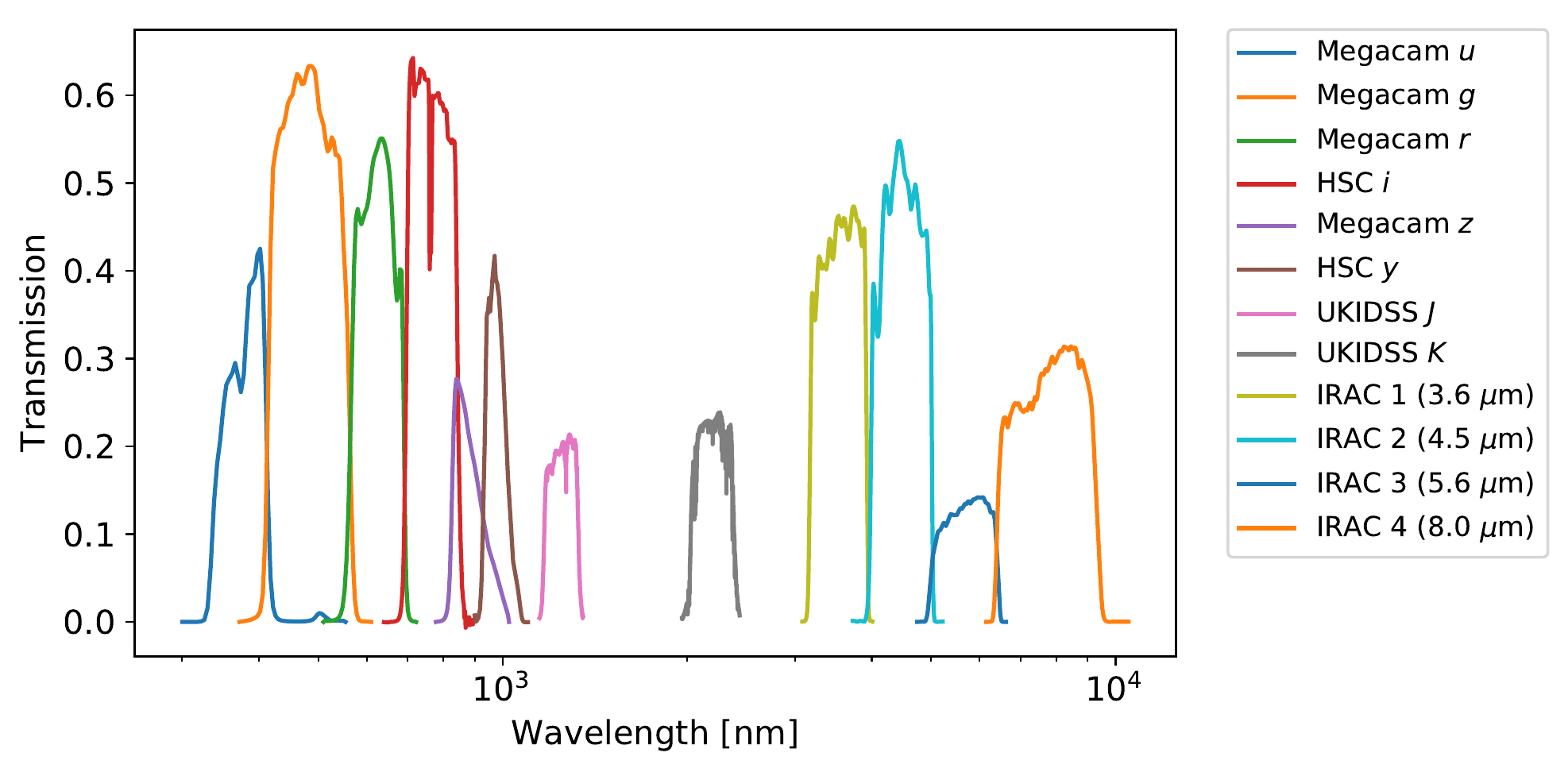}
\caption{\label{fig:filters} Filter transmission curves for a sample of filters present on the example field ELAIS-N1. Some of the optical bands have multiple measurements from different instruments with similar filters. We only show one for each band to aid clarity. These responses include quantum efficiency of the camera and atmospheric extinction as measured by the respective telescopes. Wavelength coverage varies significantly across the HELP fields. Figures for each individual field are provided in the notebooks.}
\end{figure*}

\begin{figure*}
\centering
\includegraphics[width=1.0\textwidth]{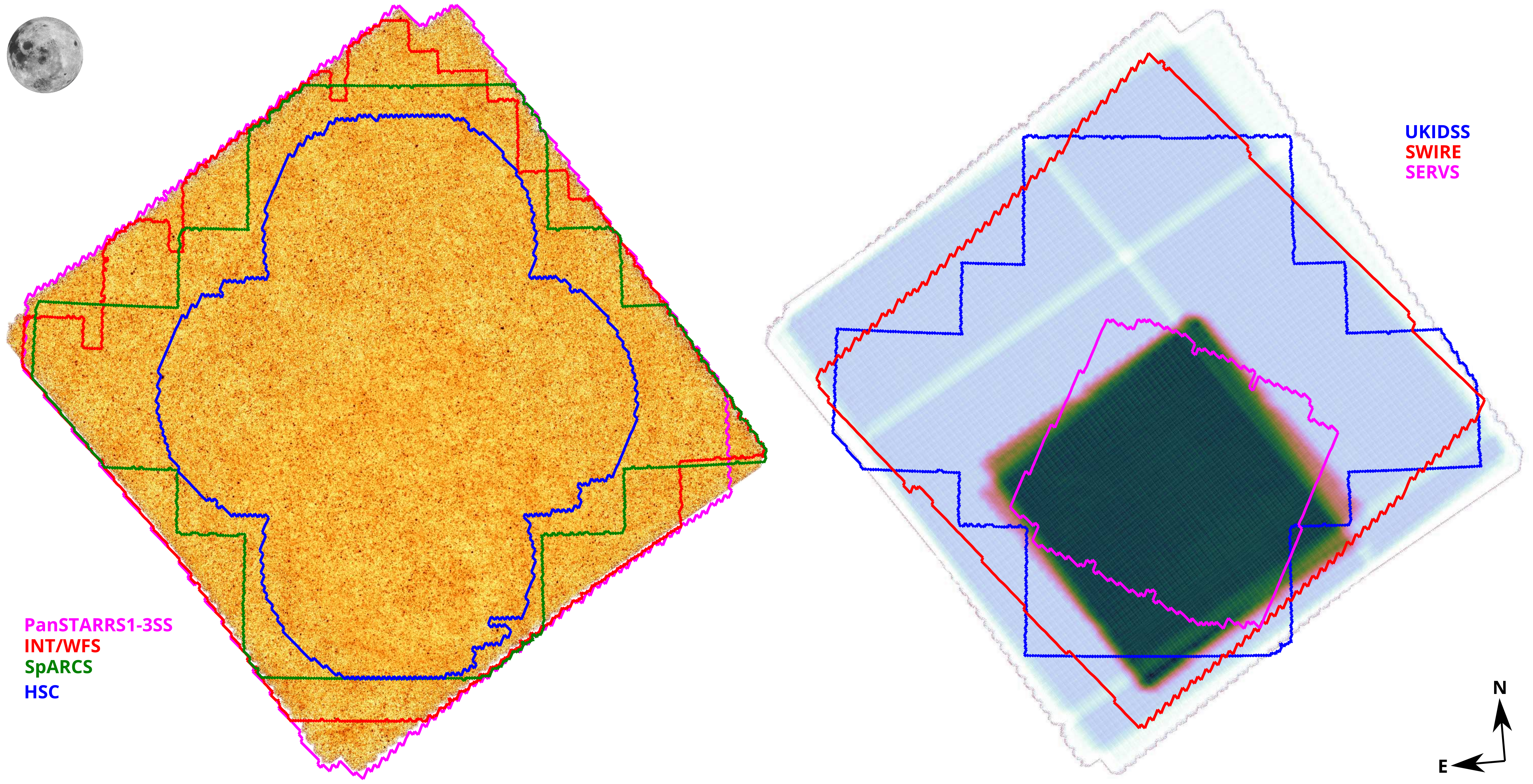}
\caption{\label{fig:coverage} Coverage of the various pristine catalogues on the example field ELAIS-N1 with Moon shown for scale. The image on the left is the SPIRE 250~$\mu$m map with the optical survey regions overlaid. The image on the right shows the SPIRE 250~$\mu$m error map with the near and mid-infrared survey regions overlaid. }
\end{figure*}

Table~\ref{table:pristine} shows all the catalogues that are available on the ELAIS-N1 field. The spectral responses of the bands available in these surveys are shown in figure~\ref{fig:filters}. Figure~\ref{fig:coverage} shows the various observation coverages overlaid on the SPIRE \( 250\ \mu \)m map\footnote{We display SPIRE here because the motivation for producing the database was to provide relevant ancillary data to the \textit{Herschel} images; for our fields, PACS data has generally the same coverage as it was acquired in parallel mode.} and corresponding variance map. 

\subsubsection{Optical data}
Observations from the Isaac Newton Telescope/Wide Field Camera (INT/WFC) cover 93\% of the ELAIS-N1 field \citep{Gonzales-Solares:2011}. This survey comprises $u,g,r,i,z$ band imaging with magnitude limits in $u$, $g$, $r$, $i$, $z$ of 23.9, 24.5, 24.0, 23.3, 22.0 respectively (AB, 5$\sigma$ point source).

The Subaru Telescope/Hyper-Suprime-Cam Strategic Program Catalogues (HSC-SSP) wide area survey covers 57\% of the ELAIS-N1 field \citep{Aihara:2018}. The survey contains imaging in five broad bands ($g,r,i,z,y$). with a 5$\sigma$ AB point-source depth of $r\approx 26$.

The \textit{Spitzer} Adaptation of the Red-sequence Cluster Survey (SpARCS) contains fluxes from the Canada France Hawaii Telescope's MegaCam instrument in $ugryz$ \citep{Tudorica:2017} down to a mean AB 5$\sigma$ $z$ depth of 24. 

\subsubsection{Near-infrared data}
The UK Infrared Telescope Deep Sky Survey - Deep Extragalactic Survey (UKIDSS-DXS) provides $J$ and $K$ band data to a 5$\sigma$ point source AB depth of $K=21.0$ \citep{Lawrence:2007, Swinbank:2017}.

\subsubsection{Mid-infrared data}
There are two separate surveys providing mid-infrared fluxes from IRAC on the \textit{Spitzer} Space Telescope. The \textit{Spitzer} Extragalactic Representative Volume Survey \citep[SERVS;][]{Mauduit:2012} provides mid-infrared fluxes to a greater depth, 5$\sigma$ AB point-source depth of $\approx 23$, over a smaller area of 2~deg.$^2$ The \textit{Spitzer} Wide InfraRed Extragalatic survey \citep[SWIRE;][]{Lonsdale:2003} provides fluxes in all four IRAC bands over 9.65~deg.$^2$ but to 5$\sigma$ AB point-source depth of $\approx 22$. In our work we use the \textit{Spitzer} data fusion products for SERVS and SWIRE as presented in \cite{Vaccari:2015}.

\begin{table*}
\scriptsize
\centering
\caption{Overview of data included on the ELAIS-N1 field. We chose the deepest public data available. Shallower data such as SDSS and 2MASS are not included because they don't provide useful extra data compared to deeper surveys. SDSS indexes are however included in order to facilitate quick identification of SDSS objects. Coverage is the percentage of the full \textit{Herschel} field observed by the given survey. A summary of the data used across all HELP fields is given in table~\ref{table:full-pristine}.}
\label{table:pristine}
\begin{tabular}{|l|l|l| l}
Input survey name                                                         & Coverage   & bands & reference\\
\hline
Isaac Newton Telescope / Wide Field Camera (INT-WFC)                      & 93\%       & $u, g, r, i, z$ & \cite{Gonzales-Solares:2011}\\
UKIRT Infrared Deep Sky Survey / Deep Extragalactic Survey (UKIDSS/DXS)   & 65\%       & $J, K$ & \cite{Swinbank:2017}\\
Hyper Suprime-Cam Subaru Strategic Program Catalogues (HSC-SSP)           & 57\%       & $g, r, i, z, y,$ $N$921, $N$816 & \cite{Aihara:2018}\\
Pan-STARRS1 - 3pi Steradian Survey (3SS) data                             & 100\%      & $g, r, i, z, y$ & \cite{Chambers:2016}\\
\textit{Spitzer} Adaptation of the Red-sequence Cluster Survey (SpARCS)   & 81\%       & $u, g, r, z$ & \cite{Tudorica:2017}\\
\textit{Spitzer} Data Fusion (SERVS)                                      & 20\%       & IRAC i1, i2 & \cite{Vaccari:2015}\\
\textit{Spitzer} Data Fusion (SWIRE)                                      & 73\%       & IRAC i1, i2, i3, i4 & \cite{Vaccari:2015} \\
\end{tabular}
\end{table*}

\begin{figure}
    \centering
\includegraphics[width=0.5\textwidth]{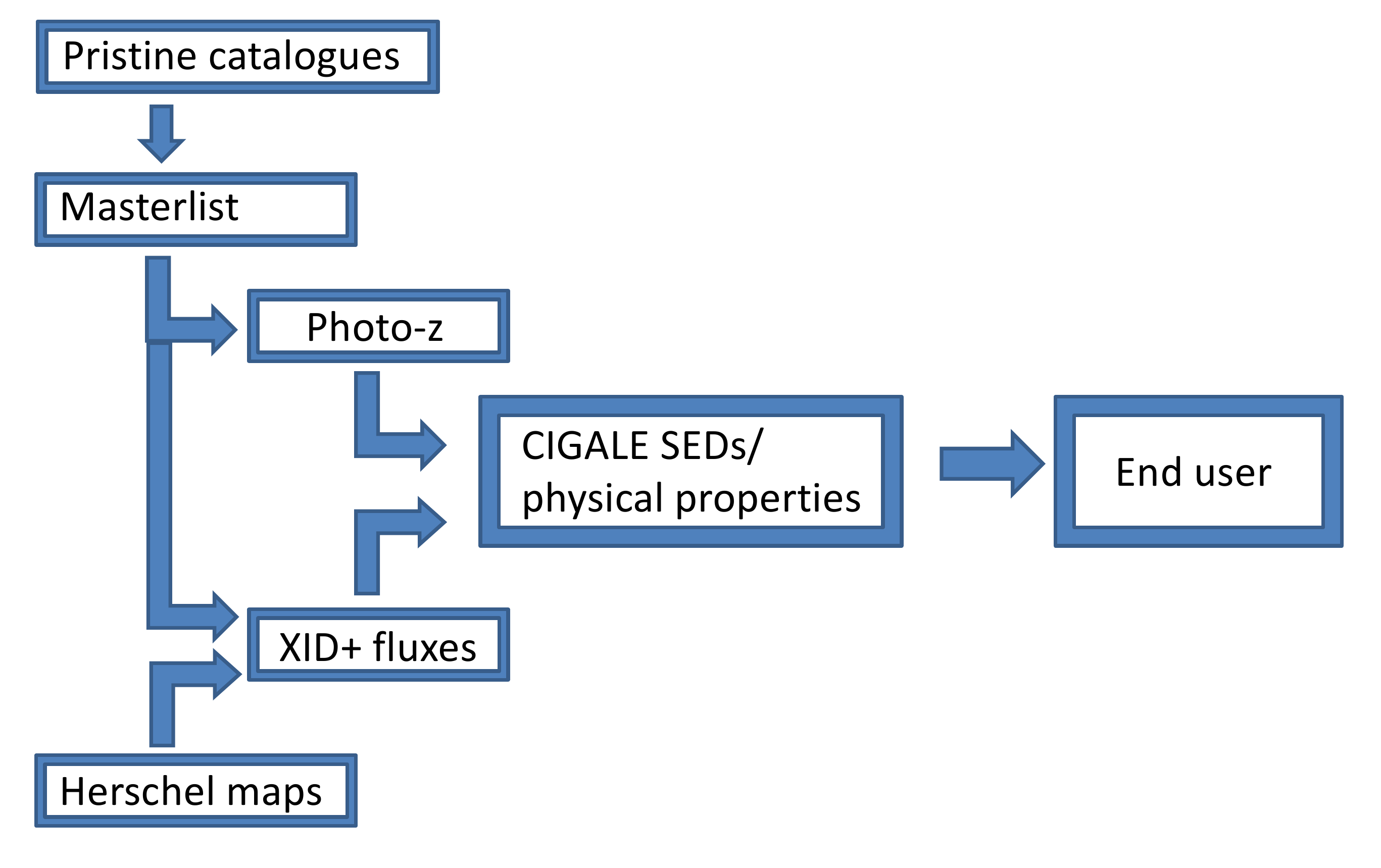}
   \caption{Schematic diagram of the full HELP pipeline. This paper concerns the production of the \emph{masterlist} which impacts all the later data processing.}
    \label{fig:schematic}
\end{figure}

Our pipeline starts with {\em pristine} catalogues provided by independent survey teams; standardises these to produce \remove{consistent format}\add{consistently formatted} {\em added value} catalogues; and merges these together to produce  multi-wavelength {\em masterlist} catalogues. These then feed in to later stages of the HELP pipeline. An overview of the full HELP pipeline is shown in figure~\ref{fig:schematic}.

\section{The \emph{masterlist} pipeline}
\label{sec:pipeline}

The entire pipeline is written in \texttt{PYTHON} making extensive use of \texttt{ASTROPY} for the cross matching \citep{Price-Whelan:2018}. Appendix~\ref{sec:data_access} gives details of how to access and run the code including all the code used to produce the figures presented here.

In addition to processing the data the pipeline performs quality  analysis and produces diagnostic plots that helped us identify errors or misunderstandings of the pristine data (e.g. Vega magnitudes reported as AB)  and provide a useful assurance for the user. These are discussed more in Section \ref{sec:validation}.

\subsection{Pristine catalogues}
We collect {\em pristine} catalogues from public data repositories, usually as delivered by the original survey teams. We provide these original tables such that one can return to the original data as it was before any corrections or conversions were carried out. This is especially useful when a particular survey contains extra information not present on all surveys and so not propagated through our pipeline such as morphology information or the results of modelling.

\subsection{Value-added catalogue preparation}

The first stage required to produce the \emph{masterlist} is to standardise the individual surveys. They must be converted into a format with consistent metadata; column headings, units, and column descriptions. Data are also set to the same astrometric reference frame and flux and magnitudes are converted to our standard of $\mu$Jy and AB magnitude respectively.

\subsubsection{Standardising fluxes and magnitudes}

In the final catalogue we provide both fluxes and magnitudes. This is done partly to make the catalogues more user friendly as both are still widely used and partly because we want to retain all information from the initial catalogues that is provided. We convert any Vega magnitudes to AB, and provide fluxes in units of $\mu$Jy.

We record both total and aperture fluxes as these have different scientific uses. The aperture fluxes are used to compute photometric redshifts (see \cite{Duncan:2018, Duncan:2017} for an overview of the HELP photometric redshift pipeline which utilises the Easy and Accurate Z(photometric redshifts) from Yale code \citep[\texttt{EAZY;}][]{Brammer:2008}). Total fluxes are used to fit the spectral energy distributions (SEDs) using the method presented in \citet{Malek:2018}, which uses the Code Investigating GALaxy Emission \citep[\texttt{CIGALE};]{Noll:2009, Roehlly:2013, Boquien:2019}. For point sources total and aperture fluxes should be the same but diverge for extended objects.
 
Total magnitudes are either Kron magnitudes or the SExtractor AUTO magnitudes \citep{Bertin:1996} chosen in that order if both are available. 2 arcsec diameter aperture magnitudes are calculated and corrected if necessary from the closest aperture to 2 arcsec. 2 arcsec is chosen as it provides a good compromise between high signal to noise and capturing a significant fraction of the total flux for a typical optical point spread function, while avoiding blended nearby objects. It is also the most commonly available aperture and consistency is a fundamental aim. Nevertheless, the choice of 2 arcsec will not be ideal in all situations. At low redshift it will yield lower signal to noise than a larger aperture for faint extended objects. It will impact selection effects and biases, through to the calculation of redshifts. Any use of these samples will have to account for this choice of total and aperture flux properties.

\subsubsection{Aperture correction}
\label{sec:aperture_correction}

\begin{figure}
\centering
\includegraphics[width=0.4\textwidth]{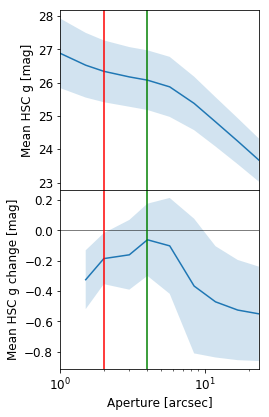}
\caption{\label{fig:aperture} Top: Hyper-Suprime-Cam uncorrected $g$-band aperture magnitude as a function of aperture size on the example field ELAIS-N1. Bottom: change in magnitude from previous aperture for point source objects. Errors are shown by the shaded region. Both figures represent the same set of objects. The vertical red line to the left shows the 2 arcsec value that is used by HELP for all surveys (or nearest available). The vertical green line to the right shows the target aperture used here (4 arcsec) to compute the average aperture correction to be applied to all objects. The curve starts to level off before showing monotonically decreasing behaviour due to contaminating background sources. }
\end{figure}

\begin{figure}
\centering
\includegraphics[width=0.4\textwidth]{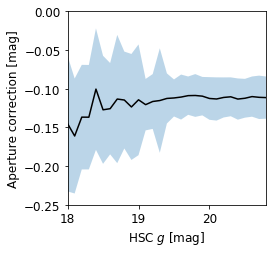}
\caption{\label{fig:aperture_cor} HSC aperture correction (difference between target, 4 arcsec, aperture magnitude and 2 arcsec aperture magnitude) as a function of $g$ band magnitude on the example field ELAIS-N1. The shaded region shows the 25th to 75th percentiles. We check it is relatively stable and take an average over the magnitude range where the correction is level; above the high corrections in very bright objects, and apply this to all objects. For the above graph point sources between 19 and 21 have a constant aperture correction which should be representative of point like galaxies at higher magnitudes. At the bright end, the sample may become subject to small number statistics and the divergence from the constant value seen between 17.5 mag and 20.0 mag may be due to point spread function features or saturation which are negligible in faint sources.}
\end{figure}

We provide aperture magnitudes in a standard 2 arcsec aperture. We provide `corrected' aperture magnitudes. The correction is to account for the fact that a 2 arcsec aperture will not encapsulate the full point spread function. The fluxes are therefore divided by the proportion of flux from a point source that is captured in the aperture. This means that for a point source they should be equal to the total magnitude. However, for extended sources they will have higher aperture magnitude than total magnitude. If these corrected aperture magnitudes are not provided by the original survey, we calculate an aperture correction by using curve of growth analysis \citep{Stetson:1990}. Figure~\ref{fig:aperture} shows how the parameters for the corrections are calculated for a typical example, HSC. This correction was applied to SpARCS and HSC data on the example field ELAIS-N1 and all surveys on other fields where corrected aperture fluxes are not provided. We look at the mean object flux in the aperture as a function of the aperture size. For a perfect zero background image in an empty field this would level off to a constant when the largest aperture included the entire point spread function. In actuality the averages start to grow as the large apertures start to include other background sources. We thus look for the point where the flux has started to level off but before it starts growing linearly with area due to unrelated sources and use that point to set a target aperture which is used to measure the ratio of the total flux for a point source to that captured in the 2 arcsec aperture. This correction, is then uniformly applied to every aperture flux. Figure~\ref{fig:aperture_cor} shows the correction as a function of magnitude. We use these figures to check that there is no significant magnitude dependence or high noise.

Fluxes and magnitudes are not corrected for galactic extinction. However, we include an E(B-V) column which can be used to apply a correction. This is done using the  the E(B-V) values from the \citet{Schlegel:1998} dust map and a scaling of 0.86 is applied to the values to reflect the re-calibration by \citet{Schlafly:2011}.\footnote{ Dust map taken from \url{https://github.com/kbarbary/sfdmap}.}

\subsubsection{Removing duplicate objects}

Some pristine catalogues included objects extracted from overlapping image tiles. This can lead to duplicates. We identify duplicates as objects within 0.4 arcsec of one and another in the same catalogue. This threshold is intended to catch most true duplicates, without excluding close pairs of intrinsically different objects. We remove the object with the highest error and flag the remaining object as having been ``cleaned''. This procedure will be influenced by the deblending algorithm used in the production of the pristine catalogue and may remove close pairs that have been correctly deblended in highly resolved catalogues. For our main aim of deblending \emph{Herschel} objects these offsets are negligible compared to the \emph{Herschel} beam but will influence photometric redshifts and SED fitting. In the deepest and highest resolution \emph{Hubble} CANDELS fields this method led to the removal of thirty objects out of over one hundred thousand.

\subsubsection{Astrometry correction}

\begin{figure}
\centering
\includegraphics[width=0.47\textwidth]{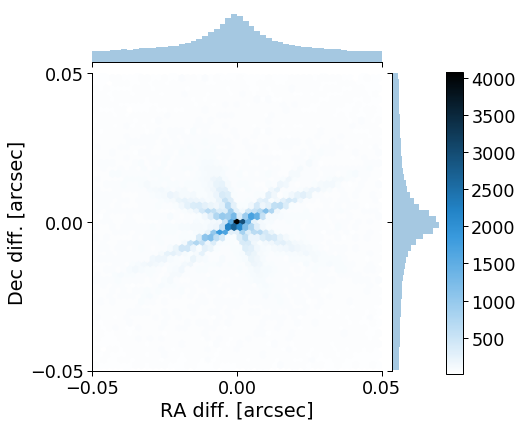}
\caption{\label{fig:astrometry1}Offsets of \emph{masterlist} positions to \emph{Gaia} reference for  objects with \emph{Gaia} $g$ magnitude larger than 18. }
\end{figure}

We convert all astrometry to the \emph{Gaia} DR1 \citep{Brown:2016} astrometric reference frame. We estimate the mean offset by performing a positional cross match with a 0.6 arcsec radius.  Offset plots are produced which can be inspected to diagnose anomalies such as significantly non-Gaussian errors. Figure~\ref{fig:astrometry1} shows the offsets of \emph{masterlist} objects compared to \emph{Gaia} objects with $g$ magnitude greater than 18. We only use the fainter objects because the bright stars often cause a large number of artefacts in the galaxy imaging surveys. The spikes in the figure and the heavy tails in the distributions are due to these. Nevertheless, the distribution is centred and tight.

\begin{figure}
\centering
\includegraphics[width=0.43\textwidth]{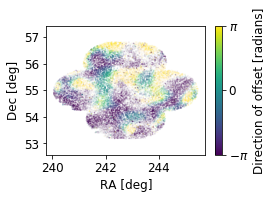}
\caption{\label{fig:astrometry2} Direction of residual astrometric offsets between HSC positions and the \emph{Gaia} reference frame, after correction for mean offset, for the example field ELAIS-N1. Colour indicates the direction and intensity indicates the size of residual offset.}
\end{figure}

We also produce a map showing the average direction of the offset for every field and every input catalogue in the diagnostic notebooks. This can reveal regular shaped regions which could result from individual tiles which should be corrected separately. Figure \ref{fig:astrometry2} gives an overview of astrometry offsets for HSC data before correction on the example field ELAIS-N1. These broadly Gaussian errors are typical of all the data sets but automatically plotted for every pristine catalogue to check for errors or inconsistencies.

We also take a unique identifier from the pristine catalogue such that any object from the final catalogue is associated with the original data set via a cross identification table and our internal HELP ID. This means that any additional information such as morphological metrics or flags for object type beyond stellarity are still available. It also means that a user with experience with one of the pristine data sets can quickly find corresponding HELP IDs for target objects to retrieve HELP data products.

\subsection{Merging the value-added catalogues}

Once all the original data has been standardised, the catalogues are merged to produce a single \emph{masterlist}. The \emph{masterlist} provides the fundamental objects fed in to all subsequent parts of the HELP workflow. 

Combining data with a positional cross match clearly comes with some limitations compared to matched aperture photometry from homogenised imaging. These include the possibility of mis-associations, and differences in selection criteria meaning selection effects are difficult to model. In the following section we investigate the performance of the data and include some metrics to aid the user in characterising the quality of their sample.

\subsubsection{The positional cross matching}

After all the pristine catalogues have been standardised we merge them together into the overall list of objects -- the \emph{``masterlist''}. This is done non-destructively: a cross-identification table can be used to return to the original tables meaning no information is discarded. 

\begin{figure}
\centering
\includegraphics[width=0.4\textwidth]{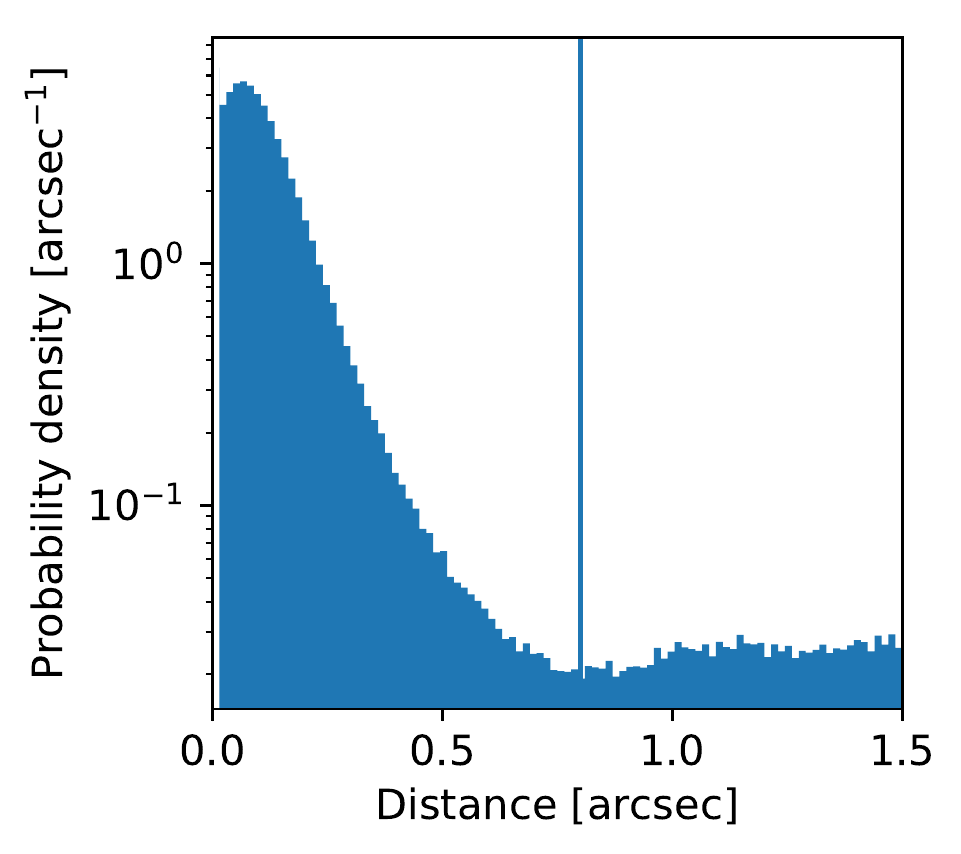}
\caption{\label{fig:crossmatch}Number density of objects from the \emph{masterlist} within a given distance from a \emph{Gaia} object with $g$ magnitude above 18. For bright stars there are large numbers of artefacts. This figure demonstrates that 0.8 arcsec is effective as a standard cross match distance. Taking all objects within 0.8 arcsec will include the majority of true matches (peak at left) and a minority of background erroneous matches (linear slope). This value is chosen for every catalogue based on the minimum in the above figure. For the majority of catalogues 0.8 arcsec (vertical line on plot) is appropriate.}
\end{figure}

We start with the highest resolution optical data and add in subsequent surveys in order of increasing positional error. We first plot the number of pairs as a function of separation. We use this to determine a maximum cross match radius. This radius varies from data set to data set but is typically chosen to be 0.8 arcsec. The threshold is chosen to capture the majority of the true associations shown by the initial bulge in figure~\ref{fig:crossmatch}. Figure~\ref{fig:crossmatch} shows the offsets of \emph{masterlist} positions with respect to \emph{Gaia}, which should reflect positional errors in general, and shows that most matches will be within 0.8 arcsec.

\begin{figure}
\centering
\includegraphics[width=0.4\textwidth]{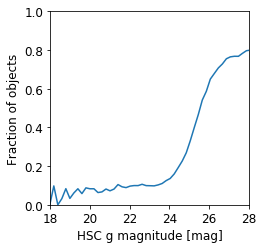}
\caption{\label{fig:merged_unmatched} Fraction of HSC objects which are flagged as merged (associated to multiple other objects) but rejected as a counterpart based on the closest match criteria as a function of HSC $g$ magnitude. These objects are unmatched and only have HSC fluxes. We expect the faintest HSC objects to be unmatched since HSC is often the deepest survey available over most of the area it exists.}
\end{figure}

Where an object has multiple associations both are included, and flagged as possibly spurious objects, but the closest match by angular separation is associated. Two percent of objects across all HELP fields have multiple associations. Magnitudes were not used as a further matching criteria. These can be included as a further criteria by selecting objects that are flagged as merged, cross matching to each other to associate merged objects and replacing unmatched objects where the magnitudes are closer. Future data releases may use a more sophisticated cross-matching or homogenised imaging and forced photometry. Tight correlations between fluxes in similar bands from different instruments show this method produces accurate cross matches (see figure~\ref{fig:mag_compare}). Figure~\ref{fig:merged_unmatched} shows the fraction of these merged objects which are unmatched as a function of magnitude. It increases rapidly with the faintest objects, which are most likely to be artifacts or undetected in another survey. The majority of objects flagged as merged will be these unmatched faint objects or the matched pair that were within 0.8 arcsec.

\subsubsection{Summary of the \remove{master-list}\add{\emph{masterlist}} detections} 
A summary of the numbers of objects in each field in the \emph{masterlist} is given in Table~\ref{table:fields}. Table~\ref{table:en1_summary} gives an overview of the numbers in the final merged catalogue across all HELP in relation to the selection criteria of objects for subsequent HELP processing.

\begin{table}
\caption{Summary of the HELP \emph{masterlist} catalogue in each field.}

\label{table:fields}
\begin{tabular}{l l r r }
Field id & HELP field name   & Objects  & Area (deg.$^2$) \\
\hline
1    &  AKARI-NEP         & 531 746    & 9.2   \\
2    &  AKARI-SEP         & 844 172    & 8.7   \\
3    &  Bootes                & 3 367 490  & 11    \\
4    &  CDFS-SWIRE        & 2 171 051  & 13    \\
5    &  COSMOS            & 2 599 374  & 5.1   \\
6    &  EGS               & 1 412 613  & 3.6   \\
7    &  ELAIS-N1          & 4 026 292  & 14    \\
8    &  ELAIS-N2          & 1 783 240  & 9.2   \\
9    &  ELAIS-S1          & 1 655 564  & 9.0   \\
10   &  GAMA-09                & 12 937 982 & 62    \\
11   &  GAMA-12                 & 12 369 415 & 63    \\
12   &  GAMA-15                & 14 232 880 & 62    \\
13   &  HDF-N                     & 130 679    & 0.67  \\
14   &  Herschel-Stripe-82  & 50 196 455 & 363   \\
15   &  Lockman-SWIRE     & 4 366 298  & 22    \\
16   &  HATLAS-NGP        & 6 759 591  & 178   \\
17   &  SA13              & 9 799      & 0.27  \\
18   &  HATLAS-SGP        & 29 790 690 & 295   \\
19   &  SPIRE-NEP         & 2 674      & 0.13  \\
20   &  SSDF              & 12 661 903 & 111   \\
21   &  xFLS              & 977 148    & 7.4   \\
22   &  XMM-13hr          & 38 629     & 0.76  \\
23   &  XMM-LSS           & 8 704 751  & 22    \\
\hline
Total:             &      & 171 570 436 & 1270 \\

\end{tabular}
\end{table}

\begin{table}
\centering
\caption{Overview of object detections across all of HELP. The main list of {\sc xid+} prior objects must be IRAC 1 detected and satisfy one of either having at least two optical detections, or at least 2 near-infrared detections. This criteria is only applied on areas observed by IRAC 1. On other areas an SED modelling method is used and will be discussed in Oliver et al. (in prep).}
\label{table:en1_summary}
\begin{tabular}{ l r  }
\hline
Total number of objects                            & 171 570 436 \\
\hline
Observed in all wavelength regimes                 &  56 618 577 \\
$\ge2$ optical detections                          & 118 750 890 \\
$\ge2$ near-infrared detections                    &  26 499 180 \\
$\ge2$ optical and $\ge2$ near-infrared detections &  22 065 144 \\
{\sc xid+} prior objects                           &  11 594 158 \\
\hline

\end{tabular}
\end{table}

\subsubsection{Stellarity indices and quality control flags}
The catalogue contains flags to identify \emph{Gaia} stars, objects that have had duplicates cleaned and objects that may have a degenerate cross match pairing. 

\begin{figure}
\centering
\includegraphics[width=0.4\textwidth]{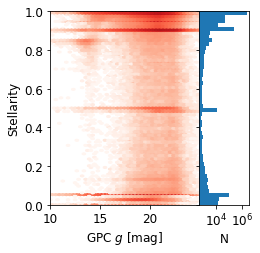}
\caption{\label{fig:stellarity} The distribution of stellarity index, which should reflect point-sourceness (definitely point like $= 1$, definitely extended $=0$) with GPC1 $g$ flux for \emph{Gaia} objects. Approximately one percent are incorrectly labeled as extended. }
\end{figure}

The stellarity is computed by taking the largest value from all the pristine catalogues. This is done on a scale where 0 represents a definitely extended object and 1 represents a definite point source. This conservative approach ensures we mark objects which are point-source like in any band. Figure~\ref{fig:stellarity} shows the distribution of stellarity values for \emph{Gaia} objects (which should all be point like and have stellarity equal to one) with GPC1 $g$ flux. This shows that approximately one percent are incorrectly labelled as extended. 

Two further flags are added to aid removal of artefacts. First we include a flag to indicate which wavelength regimes were observed at this position. This flag records whether this position on the sky was observed by any optical survey, any near-infrared survey and any mid-infrared survey. This is necessary to determine whether an absence of measurement is because the source is too faint to be detected or a given position has not been observed.

Another flag describes whether the object was detected in at least two optical bands, at least two near-infrared bands and at least two mid-infrared bands. These flags are used in later HELP products to remove artefacts.

\subsubsection{Dealing with multiple measurements}

In the \emph{masterlist} catalogue we want exactly one measurement per instrument-band. We decided to choose the lowest flux error measurement rather than combining. Our rationale is that it is not often obvious how to combine measurements rigorously and in some cases the data may not be independent. This approach means we also preserve a clear record of which survey was used for each individual object. 

An example of multiple measurements can be found on ELAIS-N1 where there are two IRAC surveys available. We investigated the distribution of errors and decided to use the deeper SERVS where it is available and the SWIRE otherwise. The depth maps described later allow us to deal with the varying depth of the resulting catalogue.

\section{Validating the final catalogue}
\label{sec:validation}
In this section we will provide details of the quality of the data along with a description of some flags used to warn the user about questionable objects or photometry values. The checks and diagnostics are automatically run following the production of the \emph{masterlist}. They were used extensively in the debugging and testing of the pipeline, consist of thousands of figures and tables and are all available on GitHub. We will show a small number of examples from ELAIS-N1 to demonstrate their utility for understanding the data.

\subsection{Photometry and stellarity}

\begin{figure}
\centering
\includegraphics[width=0.35\textwidth]{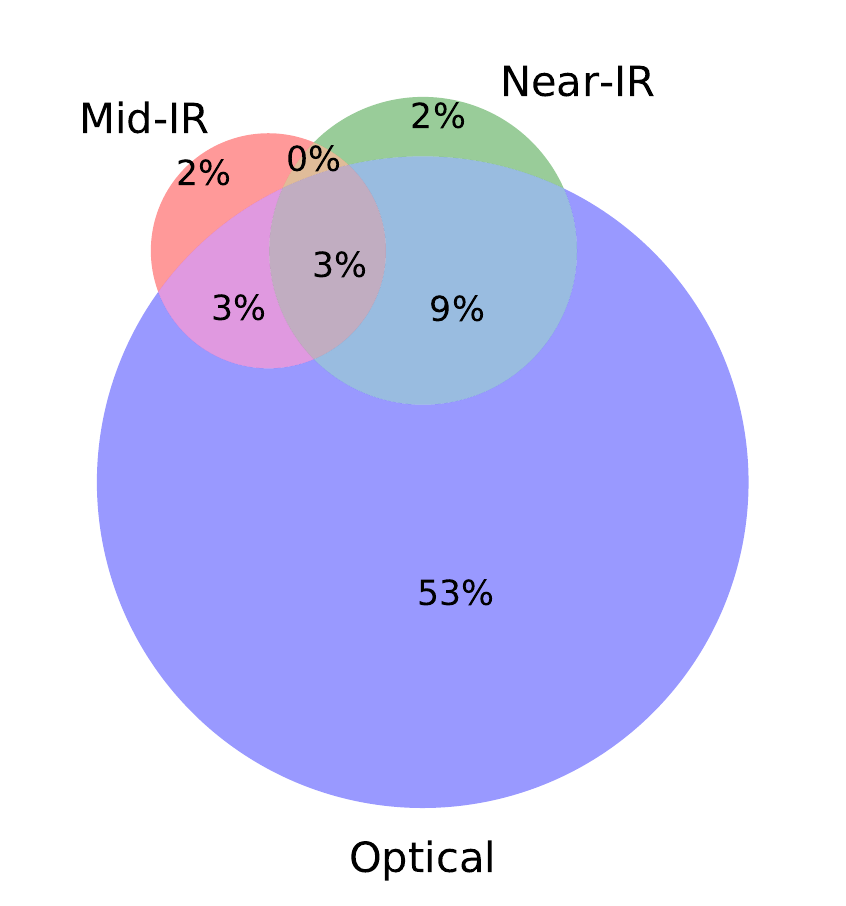}
\caption{\label{fig:detection_all} Overview of object detections across all of HELP. Optical and near-infrared objects must have at least two detections in the given wavelength regime. }
\end{figure}

\begin{figure}
\centering
\includegraphics[width=0.35\textwidth]{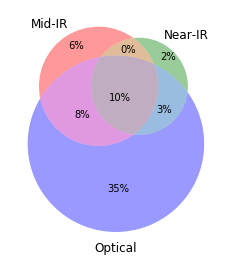}
\caption{\label{fig:detection} Overview of objects in the area, across all HELP fields, that has been observed by an optical, a near-infrared and a mid-infrared survey that have been detected in each of those regimes. The majority of objects surveyed in all wavelength regimes are only detected in an optical survey, which tend to be the deepest.}
\end{figure}

After the production of the \emph{masterlist} a number of checks and diagnostics are performed. During development of the code these facilitated the fixing of errors and validation of the data. Here, we present a discussion of each check performed and how it revealed data issues as we were constructing the software pipeline. This final stage was a crucial aspect of debugging earlier stages in the pipeline and code was developed over many iterations particularly where a full description of input data was hard to find or incomplete. We look at the numbers of objects which have detections in each combination of wavelength regimes as shown in figure~\ref{fig:detection_all} for all objects and in figure~\ref{fig:detection} for objects in regions with surveys in all wavelength regimes. The majority of objects have optical detections only due to the relative depths of each band compared to typical galaxy SEDs.

The most fundamental check is to compare magnitudes where multiple surveys give photometry for similar bands. We compare every possible combination to look for data sets that might not be compatible. On all the northern fields we have as a minimum SDSS fluxes to compare to. An example of comparing fluxes in similar bands is shown in figure~\ref{fig:mag_compare} in appendix~\ref{sec:diagnostics}. For a small number of southern fields we may not have duplicate measurements for every band so comparisons are impossible. We also compare total magnitudes to aperture magnitudes as shown in figure~\ref{fig:ap_vs_tot} to check that point sources are in strong agreement. This also functions as a check for the stellarity measure which we take as the highest value of the stellarity of each input survey. If the stellarity measures were poor, we would not see a clear distinction between point sources (stellarity > 0.7) and extended objects (stellarity < 0.7). We also plot basic number counts for every band as a further check of units and numbers of objects.

\begin{figure}
\centering
\includegraphics[width=0.35\textwidth]{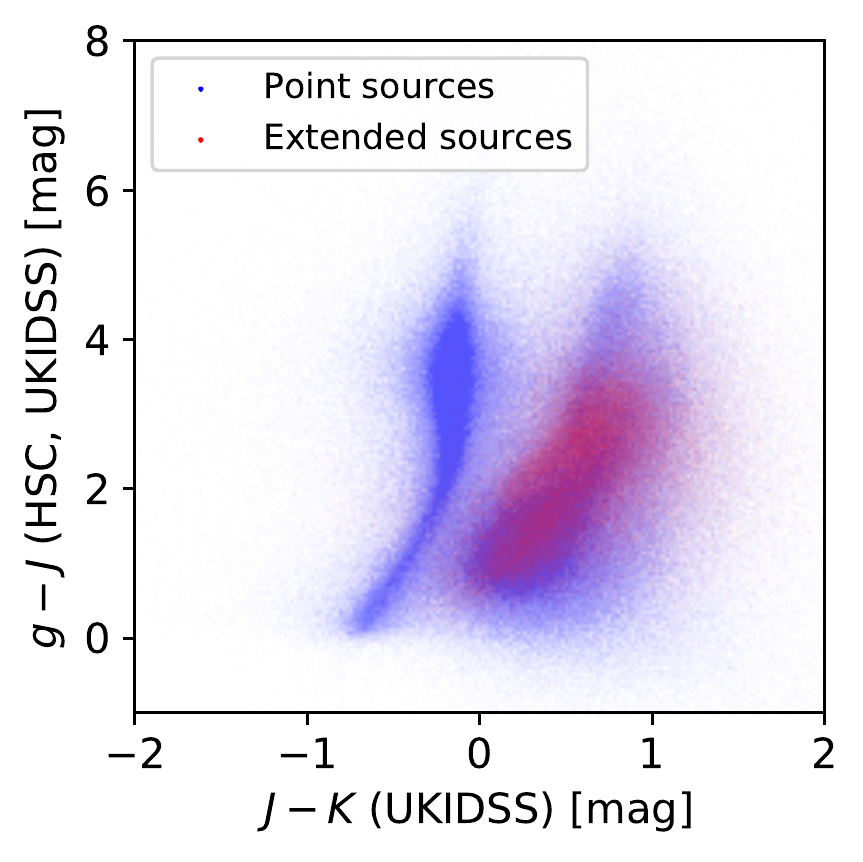}
\caption{\label{fig:colour} Colour-colour plot showing star galaxy separation for all objects with HSC $g$ and UKIDSS $K$ and $J$ fluxes across all HELP. This also serves as a check on data quality for all sets of $g$, $J$, and $K$ during catalogue production. These are plotted for a large number of colour-colour pairs in the validation section where large outliers can reveal problems with one of the bands.}
\end{figure}

In the \emph{masterlist} the stellarity parameter gives the probability that the source is extended or point-like. In order to check the accuracy of this parameter we can compare the total and aperture magnitude. As we can see in figure~\ref{fig:ap_vs_tot}, point-source objects tend to have similar aperture and total magnitudes, while for extended objects (i.e. galaxies), we get a much larger scatter. However, the stellarity parameter only gives an indication of the morphological shape of the objects. The separation between stars and quasars/galaxies is usually done using optical colours. The addition of near-infrared colours allows us to reach higher redshift. We have used up to five different colour-colour diagrams, when all wavebands were available to look at the classification of the \emph{masterlist}:

\begin{itemize}
\item Stars and galaxies can be separated using $J-K$ vs $g-i$ colours \citep{Baldry:2010}.
\item Stars can be distinguished from quasars/galaxies based on optical colours only with $g-r$ vs  $u-g$ \citep{Chiu:2007} or with a mix of optical and near-infrared colours with $g-J$ vs $J-K$ \citep{Maddox:2008} seen in figure~\ref{fig:colour}.
\item Stars, galaxies and quasars can also be identified on a $g-i$ vs $i-W1$ colour-colour diagram \citep{Tie:2017}.
\item Finally, a standard mid-infrared colour-colour diagram can be used to identify active galactic nuclei \citep{Stern:2005,Donley:2012,Lacy:2013}.
\end{itemize}

As shown in figure~\ref{fig:colour}, we observe two distributions on the $g-J$ vs $J-K$ colour-colour diagram. On one side, we have the point-source objects which correspond to the stars and on the other side we have the extended objects with some point-like sources which are galaxies and quasars respectively.

\subsection{Flags}

\begin{figure*}
\centering
\includegraphics[width=0.9\textwidth]{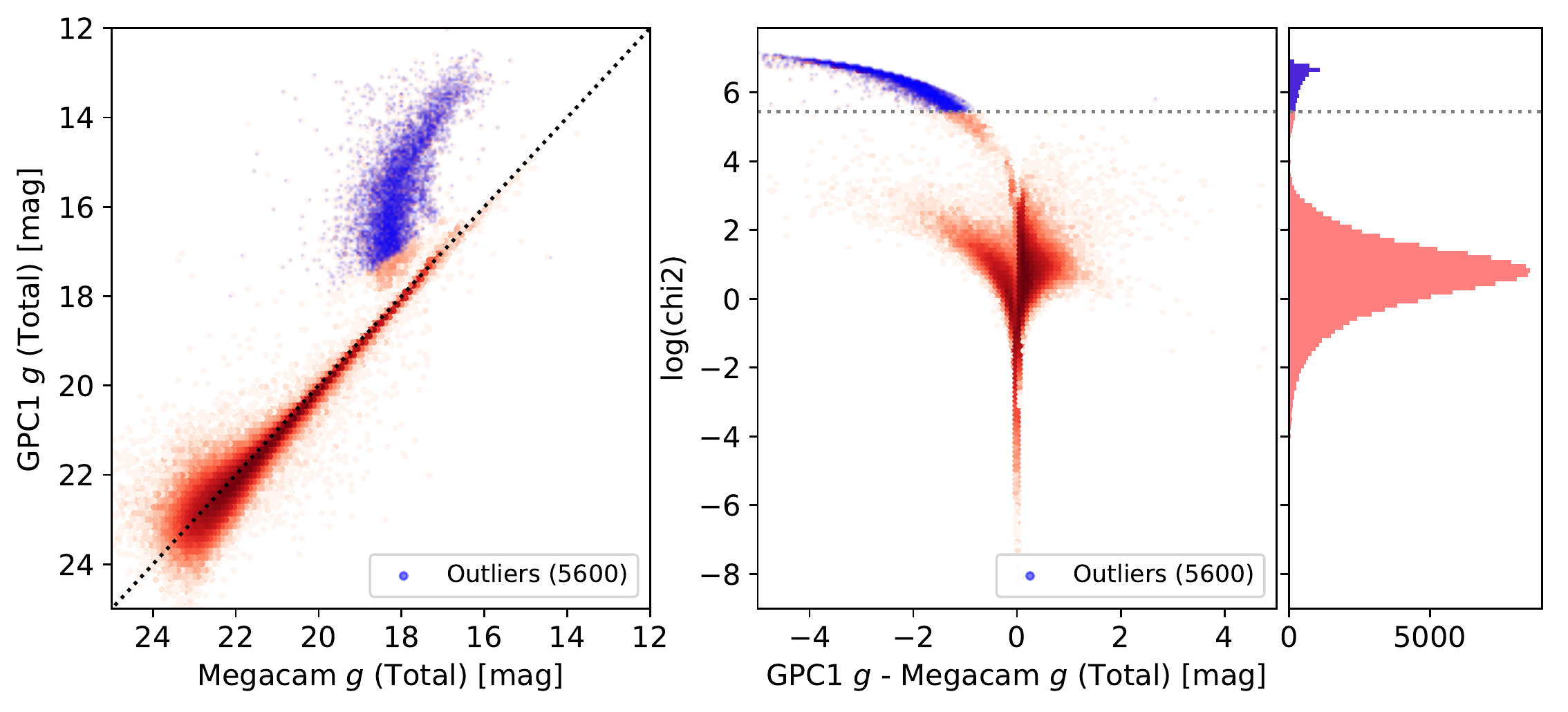}
\caption{\label{fig:outliers} Comparison between Canada France Hawaii Telescope MegaCam and PanSTARRS1 GigaPixel Camera (GPC1) $g$-band magnitudes on the example field ELAIS-N1. The blue points show those objects flagged as being outliers. They form a coherent group amongst the bright objects. We believe these are due to saturated pixels for bright objects in the SpARCS survey. The two figures to the right show the distribution of $\chi^2$ values as a function of magnitude and how the problematic fluxes form a distinct population.}
\end{figure*}

We provide some flags to aid the user in determining samples that can be used for science purposes. These are not exhaustive and merely provide some well defined metrics that will be correlated with measurement quality. We compute $\chi^2$, 

\begin{equation}
\label{eqn:chi2}
\chi^2 = \frac{(M_1 - M_2)^2}{(\sigma_1^2 + \sigma_2^2)}
\end{equation}
where $M_x$ are the magnitudes of similar bands in two different surveys and $\sigma_x$ are their corresponding errors. We flag as outliers all objects more than $5~\sigma$ from the mean. If using fluxes we recommend rerunning the notebooks with flux comparisons rather than magnitude comparisons. It can be shown that this criteria is equivalent to any pair of measurements in comparable bands having:

\begin{equation}
\label{eqn:chi2_outliers}
\chi^2 > P_\textrm{75th} + 3.2 \times (P_\textrm{75th} - P_\textrm{25th})
\end{equation}
where $P_\textrm{75th}$ and $P_\textrm{25th}$ are the 75th and 25th percentiles respectively.
Bright sources tend to have their errors underestimated with values as low as $10^{-6}$ mag, which is unrealistic. To avoid high $\chi^2$ due to these unrealistically small errors we clip the error to get a minimum value of $10^{-3}$ mag. An example of this method being applied is shown in figure~\ref{fig:outliers}.

\begin{figure}
\centering
\includegraphics[width=0.37\textwidth]{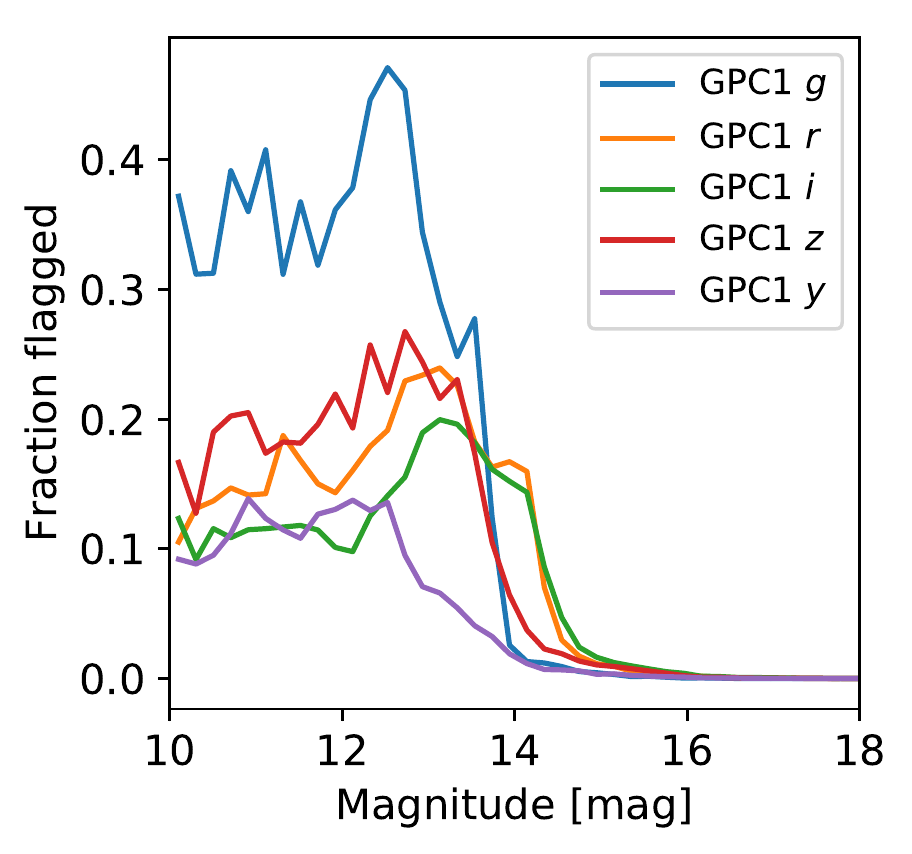}
\caption{\label{fig:flags}Fraction of flagged measurements in each of the five PanSTARRS1 GigaPixel Camera (GPC1) $grizy$ bands as a function of GPC1 $g$ magnitude across all HELP. We believe the vast majority of these flags to be due to saturated pixels in the deep DECam and MegaCam surveys, which are not optimised for bright objects. }
\end{figure}

The vast majority of flagged objects (50 thousand compared to less than ten) are due to disagreements between deep MegaCam and DECam surveys, and the wide and shallow PanSTARRS. Since in these cases there is no disagreement between PanSTARRS fluxes and numerous other surveys we believe these to be due to saturated pixels in the MegaCam and DECam surveys. We therefore recommend that for all objects with magnitude less than 16 that DECam or MegaCam fluxes are rejected in favour of fluxes from the shallower PanSTARRS if available. If this is not done then the fluxes are systematically underestimated and errors are inaccurate. Figure~\ref{fig:flags} which shows the PanSTARRS fraction of flagged objects as a function of magnitude shows how, above 18, zero objects are flagged. Since the flagging procedure is blind to which of the two fluxes has the inaccurate flux and error the PanSTARRS fluxes are also flagged in this situation.

We also implement some specific flags based on finding anomalies in the the pristine catalogues. For instance in the PanSTARRS catalogue we noticed that a number of objects had exactly the same error which is a concern. These objects are flagged. One of the major concerns with a cross matched catalogue will be the number of mis-associations. In addition to the flagging based on $\chi^2$ between measurements in similar bands we also apply a flag to objects that have multiple associations. We take the closest object as the true association and flag all objects within 0.8 arcsec as possible mis-associations.

\section{Depth maps as selection functions}
\label{sec:depth_maps}

One area of complexity is that each individual survey is subject to a different selection function; the set of criteria which determine whether a given galaxy will be in the final catalogue. Characterising this selection function from the pristine catalogues will aid the understanding of the sample of objects required for a given statistical analysis; such as using completeness to weight objects in the computation of luminosity functions \citep[e.g.][]{Loveday:2015}. As a means to describe the catalogue selection function we present `depth maps' over the field allowing one to select a sub sample according to some criteria of depth in a given band from a given instrument. For individual surveys, authors may provide measures of depth and completeness based on simulations or comparisons to deeper surveys. However, in general we may not have access to such information and a means to estimate and model survey depths and completeness from the catalogues only is highly advantageous.

To model the selection function of the catalogue we would like to know the completeness in a given band at a given location on the sky as a function of object flux. This is done using regions described by Hierarchical Equal Area Iso-Latitude pixelation of the sphere \citep[HEALPix][]{Gorski:1999}, which describes a pixelisation of the sky at varying scales or `orders'. The sphere is divided into twelve parallelograms at order zero and each subsequent order divides each parallelogram into four. The choice of order is a compromise between producing a high resolution map and having enough objects in each pixel to achieve reliable statistics. To facilitate this, and as a first step, we provide a depth at every order 10 HEALPix cell (0.003 deg.$^2$) which we take as the average error on flux on that pixel. This is done for every band for both total and aperture magnitudes. This assumes that the errors are dominated by the low flux objects, which is consistent with typical number counts but clearly there are differences between these average errors and the error on a zero flux object. Objects are typically selected according to some signal to noise criteria. Therefore errors in the total flux of an object will clearly be related to selection criteria. Nevertheless actual selection in source extraction software will depend on individual pixel measurements which are not available to us. These non-linear effects are difficult to model when we only have access to fluxes and flux errors. Since only the bands used for detection impact the selection function, the depth maps produced for bands that are not detection bands will be only correlated with the true depths to the extent that the fluxes between the bands are correlated. Further, for regions where detection is performed on a $\chi^2$ image, the depth maps produced here will only reflect correlations between flux in a given band and the $\chi^2$ value.

In the method presented here completeness is given by the probability of detecting an object given the true flux, $f_\textrm{true}$, in terms of the measured flux, $f_\textrm{measured}$ and the signal to noise cut, $n\sigma_{\textrm{mean}}$.

\begin{equation}
\label{eqn:detection}
\textrm{P(detection }| f_\textrm{true}) = \textrm{P(}f_\textrm{measured} > n\sigma_{\textrm{mean}})
\end{equation}
where $n$ is determined by the survey. This measured flux is modelled by assuming Gaussian errors on the true flux such that the completeness is given by:

\begin{equation}
\label{eqn:completeness}
\Phi(x) = \frac 1 {\sqrt{2\pi}} \int_{x}^\infty e^{-t^2/2} \, dt
\end{equation}
where $\Phi(x)$ is the completeness as a function of the measured flux in terms of a standard normal distribution such that:
\begin{equation}
x = \frac{n\sigma_{\textrm{mean}} - f_{\textrm{true}}}{\sigma_{\textrm{mean}}}
\end{equation}
and the dummy variable, $t$, is given by
\begin{equation}
t = \frac{f_{\textrm{measured}} - f_{\textrm{true}}}{\sigma_{\textrm{mean}}}
\end{equation}

We experimented with various values of $n$ between 3 and 5. Fitting $n$, by applying a single Gaussian cumulative distribution function to the SERVS number counts to recreated SWIRE yields $n \approx 5$. This is reassuring and, especially given the relative insensitivity to $n$, demonstrates that in general we can set it to 5 under the assumption that was the criteria used by the survey. 

\begin{figure}
\centering
\includegraphics[width=0.4\textwidth]{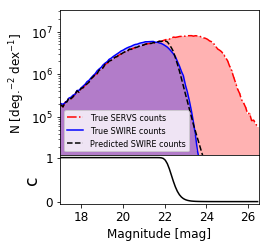}
\caption{\label{fig:completeness}IRAC $i1$ number counts on the example ELAIS-N1 field for the deeper SERVS survey and the shallower SWIRE survey in addition to predicted SWIRE counts based on our depth map selection function from SWIRE depths. Shown beneath is the SWIRE completeness, C, as calculated using the depth model in equation~\ref{eqn:completeness}. By comparing SWIRE against the deeper SERVS we see that this method produces reasonable completeness curves. }
\end{figure}

Investigations into the region covered by both SERVS and SWIRE indicate that we can successfully model the completeness of the shallower SWIRE by comparing to the `true' SERVS number counts, which should not severely suffer from incompleteness at typical SWIRE depths. Figure~\ref{fig:completeness} shows a comparison of SERVS number counts to those of SWIRE. By applying our modelled SWIRE completeness to the SERVS number counts we roughly recreate the SWIRE number counts. Moreover by freely fitting the depth criterion between $3\sigma$ and $5\sigma$ we see that completeness is relatively insensitive to the value such that assuming a $5\sigma$ cut and taking the mean errors allows the computation of reasonable completeness curves. Completeness models based on simulations which add artificial sources to real images and test retrieval will be superior to these estimates based purely on catalogues; however this method does allow one to roughly characterise depths and to set conservative values for producing magnitude limited samples.

\begin{figure}
\centering
\includegraphics[width=0.5\textwidth]{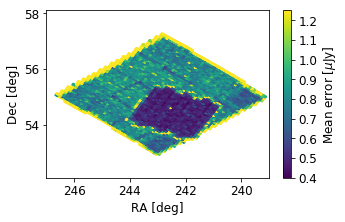}
\caption{\label{fig:depth_map} Map of the example ELAIS-N1 field showing the mean error on the IRAC i1 flux over order 10 HEALPix cells. The mean error, which can be used as a proxy for depth, makes it possible to take account of selection effects over large ares with highly inhomogeneous observation depths. The smaller and deeper SERVS area is clearly visible against the wider and shallower SWIRE area.}
\end{figure}

Figure~\ref{fig:depth_map} shows a map of the example field ELAIS-N1 with colour signifying mean error as a proxy for depth. Depths measured with this method on the COSMOS field are consistent with the depths shown in figure~3 in \citep{Laigle:2016}, which were computed from empty apertures. These depth maps are currently being used to develop a method for computing the comoving volume over which the galaxy could be detected, $V_{\textrm{max}}$, used for calculating a luminosity function that uses a different maximum redshift at which a galaxy can be detected, $z_{\textrm{max}}$, for every position on the sky.

The differences between filter transmission profiles for individual surveys will introduce systematics. Using the filter transmission curves it is possible to calculate galaxy fluxes as a function of redshift and position on the sky given which survey has observed that position and thus account for this to some extent. The decision to use 2 arcsec apertures will also feed through to the depth map values and have an effect on systematics where the completeness is far from unity. The depth maps can usefully be used in this context to set magnitude limits for the construction of samples with high completeness.

\subsection{Overview of depths and number counts}
\label{sec:depth_overview}
\begin{figure*}
\centering
\includegraphics[width=0.9\textwidth]{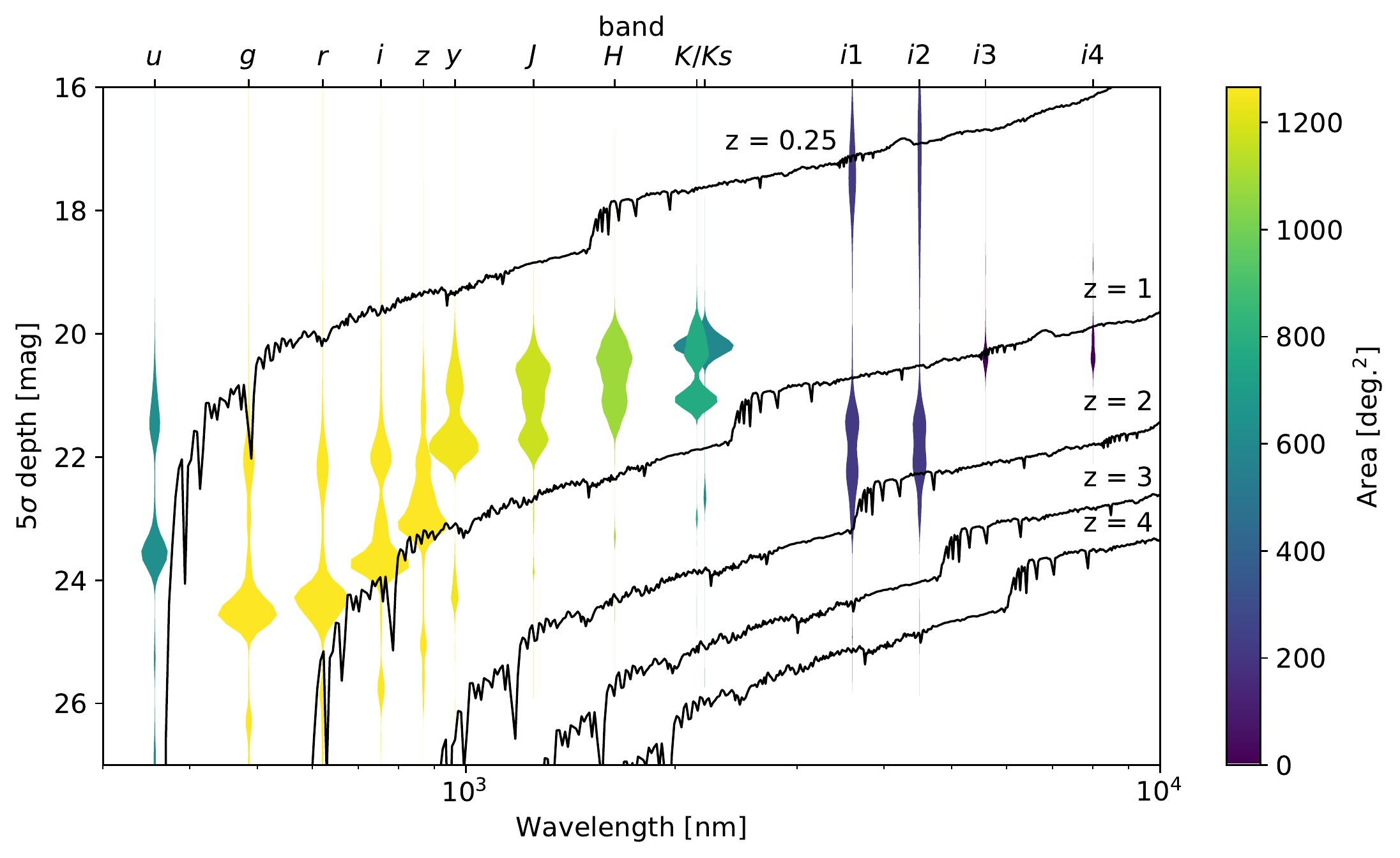}
\caption{\label{fig:bands_depths}The distribution of $5\sigma$ point-source depths in a 2 arcsec aperture for each broad band type (taking the deepest specific band available in a given HEALPix cell). The colour of each area is determined by the total area that data for that band is available. The areas of each probability density function are also weighted by the total area available for that band so that area in a given curve is proportionally related to area on the sky that a given band is available at a given depth. The combination of areas covered by either the $K$ or $K_s$ bands is the full 1270 deg.$^2$ of HELP. We also plot a typical HELP spectral energy distribution for a galaxy with star formation rate of 200~M$_\odot$/yr and a stellar mass of $10^{10}$~M$_\odot$ at various redshifts.}
\end{figure*}

\begin{figure*} 
\centering
\includegraphics[width=1.0\textwidth]{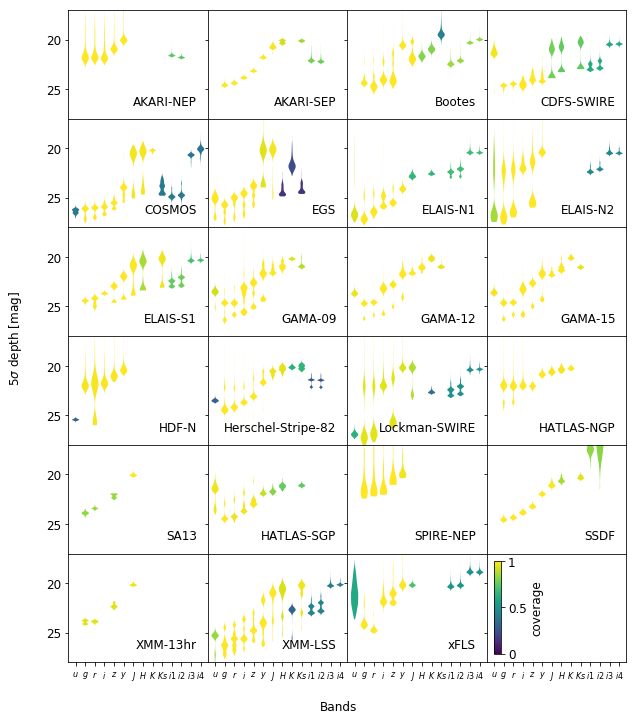}
\caption{\label{fig:fields_depths_comparison_grid} Overview of $5\sigma$ point-source aperture depths on each respective field showing the variation in coverage and depth over the fields. The violin plots show the distribution of pixel depths and thus show distributions by area. For each pixel we take the deepest specific band available. The depth maps are available for each specific band. These bands are evenly spaced for clarity and not positioned by wavelength as in figure~\ref{fig:bands_depths}. Colour represents coverage and is given as a fraction of field area. All fields have $grizy$ coverage everywhere. }
\end{figure*}

\begin{figure}
\centering
\includegraphics[width=0.4\textwidth]{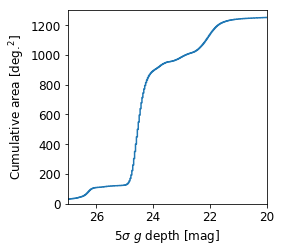}
\caption{\label{fig:g_cumulative_area_depth} The area of HELP coverage with $5\sigma$ point-source $g$ depth above a given value. The queries for generating this plot from the Virtual Observatory at SusseX (VOX) are available on GitHub. }
\end{figure}

\begin{figure}
\centering
\includegraphics[width=0.4\textwidth]{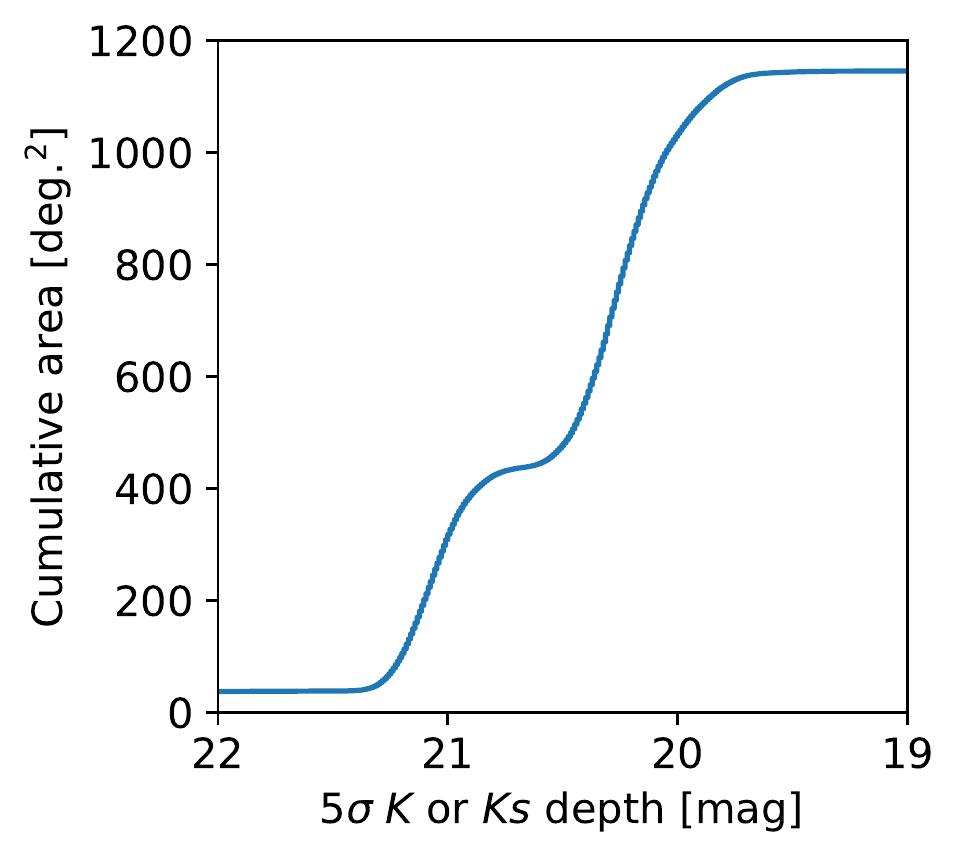}
\caption{\label{fig:K_cumulative_area_depth} The area of HELP coverage with any $5\sigma$ point-source $K$ or $Ks$ depth above a given value. The depth map product has the mean errors for all $K$ and $Ks$ bands allowing us to query the lowest value on every HEALpix cell.}
\end{figure}

\begin{figure}
\centering
\includegraphics[width=0.4\textwidth]{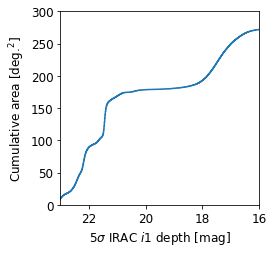}
\caption{\label{fig:IRAC_i1_cumulative_area_depth} The area of HELP coverage with $5\sigma$ point-source IRAC $i1$ band depth above a given value. IRAC $i1$ is available over 273 deg.$^2$. Building the {\sc xid+} priors on these regions requires flux prediction techniques that will be discussed in the upcoming paper by Oliver et al. Figures~\ref{fig:g_cumulative_area_depth} to \ref{fig:IRAC_i1_cumulative_area_depth} play a key role in the selection of objects for HELP processing and are a necessary component of modelling selection.}
\end{figure}

The final HELP selection function depends on optical, near-infrared and mid-infrared detections. We therefore present a number of graphs generated from the depth maps here to summarise the key depths available across HELP. We also show summaries of the number counts in $g$ and $K$ or $Ks$ to verify the depth maps reflect the numbers detected. Figure~\ref{fig:bands_depths} shows the distribution of depths over HEALPix cells for each of the main broad bands available across all of HELP. Figure~\ref{fig:fields_depths_comparison_grid} shows the distribution of depths of area on the sky for each field individually given an overview of the variation in the depths and coverage by a given type of broad band filter. We also provide individual summaries of important detection bands; figure~\ref{fig:g_cumulative_area_depth} shows the cumulative area $g$ coverage. This is made from the depth maps by taking the deepest $g$ for every HEALpix cell. Figure~\ref{fig:K_cumulative_area_depth} shows the cumulative area $K$ or $Ks$ coverage. This is made from the depth maps by taking the deepest $K$ or $Ks$ for every HEALpix cell. Together, these have coverage over the 1146 deg.$^2$ of HELP, and some coverage on all but five fields. Figure~\ref{fig:IRAC_i1_cumulative_area_depth} shows the cumulative area IRAC $i1$ coverage. The depth percentiles quoted in the abstract correspond to the distribution of depths by area covered on the sky as displayed in these figures.

We also provide cumulative depth plots for bands $g$ (figure~\ref{fig:g_cumulative_area_depth}), $K$ or $K_s$ (figure~\ref{fig:K_cumulative_area_depth}) and IRAC $i1$ (figure~\ref{fig:IRAC_i1_cumulative_area_depth}) for a more detailed overview of the depths available in these bands.

\begin{figure}
\centering
\includegraphics[width=0.4\textwidth]{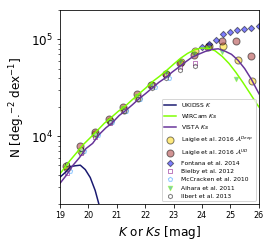}
\caption{\label{fig:numbers_K_COSMOS_comparison} $K$ or $Ks$ selected number counts for the COSMOS field. HELP numbers (this study) are shown by the lines. The points show the number counts from \citet{McCracken:2009,  Aihara:2011,   Bielby:2012, Fontana:2014, Ilbert:2015, Laigle:2016}.}
\end{figure}

\begin{figure*} 
\centering
\includegraphics[width=1.0\textwidth]{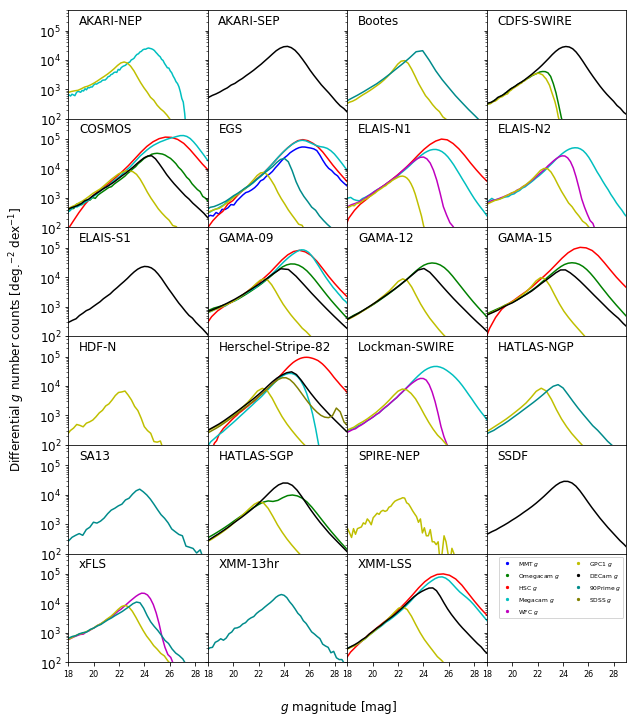}
\caption{\label{fig:numbers_g_allfields} The differential $g$ (total magnitude) number counts for the most frequently encountered wide area $g$ broad band fluxes available in the catalogue. These are computed using the area over which a given band is available on a given field. Occasionally two peaks can be seen where a given survey has areas with different depths. This effect, which corresponds to using an overestimated area for the deeper part of the survey, can be seen on Herschel-Stripe-82 and HATLAS-SGP for instance.  }
\end{figure*}

\begin{figure*} 
\centering
\includegraphics[width=1.0\textwidth]{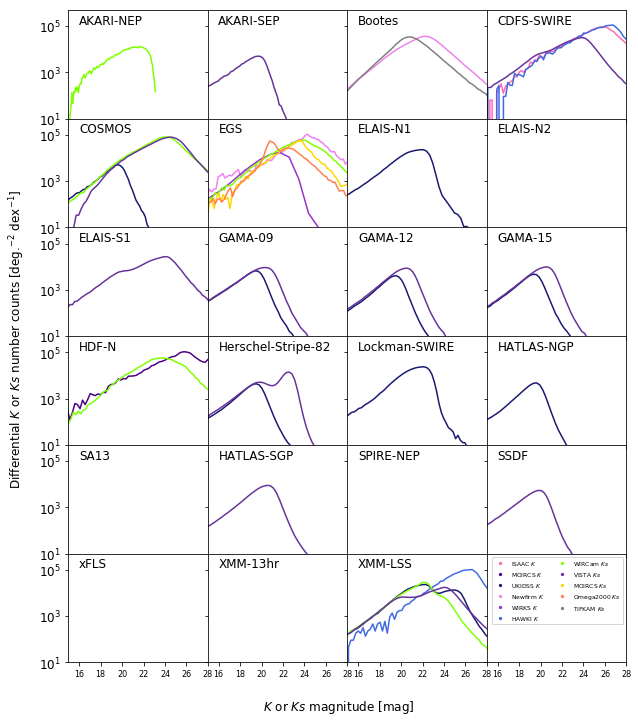}
\caption{\label{fig:numbers_K_allfields} The differential $K$, or $Ks$ (total magnitude) number counts for the most frequently encountered wide area $K$, or $Ks$ broad band fluxes available in the catalogue. Occasionally, where a survey has regions in the field to different depths there are multiple peaks. These are computed using the area over which a given band is available on a given field. There are five fields without any $K$ or $Ks$ data. These are ELAIS-N2, SA13, SPIRE-NEP, xFLS, and XMM-13hr.  }
\end{figure*}

We also provide figures showing the number counts for each field. Figure~\ref{fig:numbers_K_COSMOS_comparison} highlights the COSMOS field and compares to the counts from the following sources \citet{McCracken:2009,  Aihara:2011, Bielby:2012, Fontana:2014, Ilbert:2015, Laigle:2016}. Figures~\ref{fig:numbers_g_allfields} and \ref{fig:numbers_K_allfields} which show the differential number counts on each field for $g$ and either $K$ or $Ks$. The areas used to compute these are the total area over which a given band is available on a given field. For this reason there can be two peaks where a given survey has different areas with varying depths.

\section{Conclusions}
\label{sec:conclusions}
We have presented a new multi-wavelength catalogue across the well known and well studied extragalactic fields that were targeted by the \textit{Herschel} Space Observatory. This new catalogue will be of general use to extragalactic astronomers and forms the basis of all data products currently being produced for the \textit{Herschel} Extragalactic Legacy Project (HELP). This catalogue defines the list of objects that will comprise the first data release from the project (HELP-DR1). We have described a method for producing and testing multi-wavelength catalogues from disparate and inhomogeneous surveys for use in wide area astronomy. We discussed a number of problems involved with collating data of differing quality and production methods. The resulting catalogue will be used in various upcoming projects including physical modelling and forced photometry from low resolution \textit{Herschel} imaging.

To summarise, the paper presents:
\begin{itemize}
\item A new multi-wavelength catalogue on all HELP fields covering 1270 deg.$^2$, with astrometry corrected to the \emph{Gaia} frame and a fully documented positional cross match, that is reproducible and extensible in an open science framework.

\item A full description of the catalogue production methods that have been applied across all HELP fields. The code used to perform this reduction in addition to thousands of diagnostic plots are provided in \texttt{JUPYTER} notebooks that are available to download. We provide summary plots showing data quality across the whole HELP coverage. Equivalent plots for individual fields are available in the open access notebooks. 

\item Diagnostics and depth maps that demonstrate and measure the quality and limitations of the data in addition to flags to warn the user about possibly spurious objects. The depth maps can be used to model selection functions, facilitating the construction of a well understood sample of objects for statistical analysis.
\end{itemize}

\section*{Acknowledgements}
The research leading to these results has received funding from the European Union Seventh Framework Programme FP7/2007-2013/ under grant agreement no.607254. This publication reflects only the authors' view and the European Union is not responsible for any use that may be made of the information contained therein. We are grateful to the anonymous referee for comments that have lead to a significant improvement in the presentation of these results. Raphael Shirley acknowledges support from the Daphne Jackson Trust and from the Spanish Ministerio de Ciencia, Innovaci\'{o}n y Universidades (MICINN) under grant numbers ESP2017-86582-C4-2-R and ESP2015-65597-C4-4-R. Katarzyna Małek has been partly supported by the National Science Centre (grant UMO-2018/30/E/ST9/00082). Lucia Marchetti and Mattia Vaccari have been partially supported by the South African Department of Science and Technology (DST/CON 0134/2014). This work has made use of the COSMOS2015 catalogue \citep{Laigle:2016} based on data products from observations made with ESO Telescopes at the La Silla Paranal Observatory under ESO programme ID 179.A-2005 and on data products produced by TERAPIX and the Cambridge Astronomy Survey Unit on behalf of the UltraVISTA consortium.

\bibliographystyle{mnras} 
\interlinepenalty=10000
\bibliography{./bibliography}



\appendix

\section{Summary of input data sets and coverage}
\label{sec:appendix}

Here we include further details of the 51 surveys included in the \emph{masterlist} and their respective coverage of HELP fields. Table~\ref{table:fields} shows the HELP fields with a basic overview of the field size and number of catalogue objects. A full description of the HELP fields from the perspective of \emph{Herschel} imaging and downstream HELP data processing will be provided in the upcoming paper by Oliver et al. Table~\ref{table:full-pristine} gives an overview of all data included here. Table~\ref{table:survey_field_coverages} gives an overview of the coverages of each survey on each field including the mean depth. Table~\ref{table:band_coverages} gives an overview of the main broad band photometry available on each field including the median depths. Further details regarding the input datasets, including where they can be downloaded, pre-processing we performed and documentation regarding their units etc are provided in the database documentation on GitHub in data management unit 0 (dmu0). The code used to produce the multi-wavelength catalogues created from these is presented in dmu1. All other numbered dmu folders contain details of aspects of the HELP pipeline which are not presented here.

\begin{landscape}
\thispagestyle{empty}
\begin{table}
\tiny
\centering
\caption{Overview of data included on all HELP fields. We chose the deepest public data available. }
\label{table:full-pristine}
\begin{tabular}{l p{6cm} p{4cm} p{6cm} p{4cm} p{3cm} }

Survey id & Survey                                                                                             & Telescope                                                     & Instrument                                                       & Filters                              &   Reference                                     \\
\hline                                                                                                                                                                                                                                                                                                                        
1  & Two Micron All Sky Survey (2MASS)                                                                  & 2MASS dedicated telescopes                                    & Dedicated instrument                                             & J, H, Ks                             & \cite{Skrutskie:2006, Cutri:2003}           \\ 
2  & All-wavelength Extended Groth strip International Survey (AEGIS)                                   & Palomar Observatory                                           & WIRC                                                             & J, Ks                                & \cite{Bundy:2006}               \\ 
3  & Optical-NIR catalog of AKARI NEP Deep Field                                                        & Canada France Hawaii Telescope (CFHT)                         & MegaCam, WIRCam                                                  & u, g, r, i, z, Y, J, Ks              & \cite{Oi:2014}                  \\ 
4  & VST ATLAS SURVEY                                                                                   & VLT Survey Telescope (VST)                                    & OmegaCAM                                                         & u, g, r, i, z                        & \cite{Shanks:2015}              \\ 
5  & 3D-HST+CANDELS catalog                                                                             & Hubble Space Telescope (HST)                                  & Wide FIeld Camera 3 (WFC3)                                       & 5 HST bands (125nm--606nm)           & \cite{Skelton:2015}             \\ 
6  & Cosmic Assembly Near-IR Deep Extragalactic Legacy Survey (CANDELS)                                 & HST                                                           & WFC3 , Advanced Camera for Surveys (ACS)                         & 10 HST bands (98nm--850nm)           & \cite{Grogin:2011}              \\ 
7  & WIRcam Deep Survey (WIRDS)                                                                         & CFHT                                                          & WIRCAM                                                           & J, H, Ks                             & \cite{Bielby:2012}              \\ 
8  & Canada-France-Hawaii Telescope Legacy Survey (CFHTLS)                                              & CFHT                                                          & MegaCam                                                          & u, g, r, i, z                        & \cite{Hudelot:2012}             \\ 
9  & The CFHT Lensing Survey (CFHTLenS)                                                                 & CFHT                                                          & MegaCam                                                          & u, g, r, i, z                        & \cite{Heymans:2012, Erben:2013} \\ 
10 & Classifying Objects by Medium-Band Observations 17-filter survey (COMBO-17)                        & MPG/ESO 2.2m Telescope                                        & Wide Field Imager (WFI)                                          & UBVRI, 12 narrow bands (420nm--914nm) & \cite{Wolf:2004}                \\ 
\hline  
   & DESI Legacy Imaging Surveys:\\
11 & DECam Legacy Survey (DECaLS)                                                                       & Bok Telescope, Mayall Telescope, Blanco Telescope             & 90Prime camera, MOSAIC-3 camera, Dark Energy Camera (DECam)      & u, g, r, i, z, Y                     & \cite{Blum:2016, Dey:2019}                \\ 
12 & Beijing–Arizona Sky Survey, and the Mayall z-band Legacy Survey         & Bok Telescope, Mayall Telescope                               & 90Prime camera, MOSAIC-3 camera                                  & u, g, r, i, z, Y                     & \cite{Zou:2017, Silva:2016, Dey:2019}     \\ 
\hline  
13 & DEEP2 Galaxy Redshift Survey                                                                       & CFHT                                                          & CFH12K                                                           & B, R, I                              & \cite{Coil:2004}                \\ 
14 & Dark Energy Survey (DES)                                                                           & Blanco Telescope                                              & Dark Energy Camera (DECam)                                       & u, g, r, i, z, Y                     & \cite{Abbott:2018}              \\ 
15 & Spitzer Data Fusion Catalogues (SERVS and SWIRE)                                                   & Spitzer                                                       & IRAC, MIPS                                                       & IRAC1234, MIPS123                    & \cite{Vaccari:2015}             \\ 
16 & ESO-Spitzer Imaging extragalactic Survey (ESIS)                                                    & MPG/ESO 2.2m Telescope                                        & Wide Field Imager (WFI)                                          & B, V, R                              & \cite{Vaccari:2016}               \\ 
17 & FIREWORKS photometry of GOODS CDF-S                                                                & MPG/ESO 2.2m Telescope, HST, VLT, Spitzer                     & ESO/WFI, HST/ACS, VLT/ISAAC, Spitzer/IRAC                        & 16 bands UV to FIR                   & \cite{Wuyts:2008}               \\ 
18 & GOODS ACS Treasury program                                                                         & HST                                                           & ACS                                                                  & b, i, v, z                           & \cite{Giavalisco:2004}          \\ 
19 & Hyper Suprime-Cam Subaru Strategic Program (HSC-SSP)                                               & Subaru                                                        & Hyper Suprime-Cam (HSC)                                          & g, r, i, z, y, N921, N816            & \cite{Aihara:2018}              \\ 
20 & Hawaii Hubble Deep Field North (Hawaii-HDFN)                                                       & Mayall Telescope, Subaru, UH 2.2m Telescope                   & MOSAIC-1, Suprime-Cam, QUIRC                                     & U, B, V, R, I, z', HK'               & \cite{Capak:2004}               \\ 
21 & Infrared Bootes Imaging Survey (IBIS)                                                              & Mayall Telescope                                              & NEWFIRM                                                          & J, H, Ks                             & \cite{Jannuzi:2004}       \\ 
22 & Wide-field optical imaging on ELAIS N1, ELAIS N2, First Look Survey and Lockman Hole               & Isaac Newton Telescope (INT)                                  & Wide Field Camera (WFC)                                          & u, g, r, i, z                        & \cite{Gonzales-Solares:2011}    \\ 
23 & UV-to-FIR Analysis of Spitzer/IRAC Sources in the Extended Groth Strip                             & Multiple                                                      & Multiple                                                         & Multiple                             & \cite{Barro:2011}               \\ 
24 & Kilo-Degree Survey (KiDS)                                                                          & VLT Survey Telescope (VST)                                    & OmegaCAM                                                         & u, g, r, i                           & \cite{Jong:2013}                \\ 
25 & Spitzer First Look Survey (FLS) - NOAO Extragalactic - R                                           & Mayall Telescope                                              & MOSAIC-1                                                         & R                                    & \cite{Fadda:2004}               \\ 
26 & NOAO Deep Wide-Field Survey (NDWFS)                                                                & Mayall Telescope                                              & MOSAIC-1                                                         & Bw, R, I, Ks                         & \cite{Jannuzi:1999}             \\ 
27 & Panoramic Survey Telescope and Rapid Response System (PanSTARRS1) 3pi Steradian Survey (3SS)       & PanSTARRS1 (PS1)                                              & Gigapixel Camera 1 (GPC1)                                        & g, r, i, z, y                        & \cite{Chambers:2016}            \\ 
28 & Red Cluster Sequence Lensing Survey (RCSLenS)                                                      & CFHT                                                          & MegaPrime/MegaCam                                                & g, r, i, y, z                        & \cite{Hildebrandt:2016}         \\ 
29 & Sloan Digital Sky Survey (SDSS) - DR13                                                             & Sloan Foundation 2.5m Telescope                               & SDSS camera                                                      & u, g, r, i, z                        & \cite{Albareti:2017}            \\ 
30 & SDSS - Stripe 82                                                                                   & Sloan Foundation 2.5m Telescope                               & SDSS camera                                                      & u, g, r, i, z                        & \cite{Viero:2014}               \\ 
31 & Instituto de Astrofísica de Canarias (IAC) Stripe 82 Legacy Project                                & Sloan Foundation 2.5m Telescope                               & SDSS camera                                                      & u, g, r, i, z                        & \cite{Fliri:2016}               \\
32 & Spitzer Deep, Wide-Field Survey (SDWFS)                                                            & Spitzer                                                       & IRAC                                                             & IRAC1234                             & \cite{Ashby:2009}               \\ 
33 & Spitzer/HETDEX Exploratory Large-Area (SHELA) survey                                               & Spitzer                                                       & IRAC                                                             & IRAC12                               & \cite{Papovich:2016}            \\ 
34 & Spitzer-IRAC/MIPS Extragalactic Survey (SIMES)                                                     & Spitzer, Herschel                                             & IRAC, MIPS, SPIRE                                                & IRAC12                               & \cite{Baronchelli:2016}         \\ 
35 & SPLASH-SXDF Multi-Wavelength Photometric Catalog                                                   & Multiple                                                      & Multiple                                                         & Multiple                             & \cite{Mehta:2017}               \\ 
36 & Spitzer-South Pole Telescope Deep Field (SSDF)                                                     & Spitzer                                                       & IRAC                                                             & IRAC12                               & \cite{Ashby:2013}               \\ 
37 & Subaru/XMM-Newton Deep Survey (SXDS)                                                               & Subaru                                                        & Suprime-Cam                                                      & B, V, r, i, z                        & \cite{Furusawa:2008}            \\ 
38 & Spitzer Adaptation of the Red-sequence Cluster Survey (SpARCS)                                     & CFHT                                                          & MegaCam                                                          & u, g, r, y, z                        & \cite{Tudorica:2017}            \\ 
39 & Spitzer IRAC Equatorial Survey (SpIES)                                                             & Spitzer                                                       & IRAC                                                             & IRAC12                               & \cite{Timlin:2016}              \\ 
40 & Spitzer UKIDSS Ultra Deep Survey (SpUDS)                                                           & Spitzer                                                       & IRAC                                                             & IRAC1234                             & \cite{Dunlop:2007}              \\ 
41 & UKIRT Hemisphere Survey (UHS)                                                                      & United Kingdom Infrared Telescope (UKIRT)                     & WFCAM                                                            & J                                    & \cite{Dye:2018}                 \\ 
42 & UKIDSS Deep eXtra-galactic Survey (DXS)                                                            & UKIRT                                                         & WFCAM                                                            & J, K                                 & \cite{Swinbank:2017}            \\ 
43 & UKIDSS Large Area Survey (LAS)                                                                     & UKIRT                                                         & WFCAM                                                            & U, J, H, K                           & \cite{Lawrence:2007}            \\ 
44 & UKIDSS Ultra-Deep Survey (UDS)                                                                     & UKIRT                                                         & Wide Field Infrared Camera (WFCAM)                               & J, H, K                              & \cite{Almaini:2007}             \\ 
45 & Ultradeep Ks Imaging in the GOODS-N                                                                & CFHT                                                          & WIRCam                                                           & Ks (plus IRAC crossmatches)          & \cite{Wang:2010}                \\ 
46 & VISTA-CFHT Stripe 82 Survey (VICS82)                                                               & CFHT, VISTA                                                   & CFHT/WIRCam, VISTA/VIRCAM                                        & J, Ks                                & \cite{Geach:2017}               \\ 
47 & VIPERS Multi-Lambda Survey (MLS)                                                                   & CFHT + others                                                 & WIRCam (+ others)                                                & Ks selected multiband                & \cite{Moutard:2016}             \\ 
48 & Visible and Infrared Survey Telescope for Astronomy (VISTA) Hemisphere Survey (VHS)                & VISTA                                                         & VISTA InfraRed CAMera (VIRCAM)                                   & Y, J, H, Ks                          & \cite{McMahon:2013}             \\ 
49 & VISTA Deep Extragalactic Observations (VIDEO) Survey                                               & VISTA                                                         & VIRCAM                                                           & Y, J, H, Ks                          & \cite{Jarvis:2013}              \\ 
50 & VISTA Kilo-Degree Infrared Galaxy Survey (VIKING)                                                  & VISTA                                                         & VIRCAM                                                           & Y, J, H, Ks                          & \cite{Edge:2013}                \\ 
51 & zBootes - z-band Observations of the NOAO Deep Wide-Field Survey Bootes Field                      & Bok Telescope                                                 & 90Prime imager                                                   & z                                    & \cite{Cool:2007}                \\ 

\end{tabular}
\end{table}
\end{landscape}

\begin{landscape}
\thispagestyle{empty}
\begin{table}
\tiny
\caption{Availability on a given HELP field of each input survey. Table~\ref{table:fields} describes the id number used for each field. Survey names in there expanded form with references are given in Table~\ref{table:full-pristine} based on the survey id. }
\label{table:survey_field_coverages}
\begin{tabular}{l l l l l l l l l l l l l l l l l l l l l l l l l l l }
\hline
Survey id & Survey & area (deg.$^2$) & No. fields  & \multicolumn{23}{c}{Area of survey coverage for each field in deg.$^2$ (Use Table~\ref{table:fields} for key)}\\
\cline{5-27}
&       &                 &          &1&2&3&4&5&6&7&8&9&10&11&12&13&14&15&16&17&18&19&20&21&22&23\\
\hline
1 & 2MASS               & 1270 (100\%)    & 23  & 9.2 & 8.7 & 11 & 13 & 5.1 & 3.6 & 14 & 9.2 & 9.0 & 62 & 63 & 62 & 0.7 & 363 & 22 & 178 & 0.3 & 294 & 0.1 & 111 & 7.4 & 0.8 & 22 \\ 
2 & AEGIS               & 0.7 (0.1\%)     & 1  &     &     &     &     &     & 0.7 &     &     &     &     &     &     &     &     &     &     &     &     &     &     &     &     &     \\ 
3 & AKARI-NEP-OptNIR    & 1.1 (0.1\%)& 1  & 1.1 &     &     &     &     &     &     &     &     &     &     &     &     &     &     &     &     &     &     &     &     &     &     \\ 
4 & ATLAS               & 308 (24\%)& 2  &     &     &     & 13 &     &     &     &     &     &     &     &     &     &     &     &     &     & 295 &     &     &     &     &     \\ 
5 & CANDELS-3D-HST      & 0.3 (0.02\%)& 5  &     &     &     & 0.1 & 0.1 & 0.1 &     &     &     &     &     &     & 0.1 &     &     &     &     &     &     &     &     &     & 0.1 \\ 
6 & CANDELS             & 0.2 (0.02\%)& 4  &     &     &     & 0.1 &     & 0.1 &     &     &     &     &     &     & 0.1 &     &     &     &     &     &     &     &     &     & 0.1 \\ 
7 & CFHT-WIRDS          & 2.3 (0.2\%)& 3  &     &     &     &     & 1.0 & 0.5 &     &     &     &     &     &     &     &     &     &     &     &     &     &     &     &     & 0.8 \\ 
8 & CFHTLS              & 24 (1.9\%)& 4  &     &     &     &     & 1.1 & 3.6 &     &     &     & 4.3 &     &     &     &     &     &     &     &     &     &     &     &     & 15 \\ 
9 & CFHTLenS            & 21 (1.7\%)& 3  &     &     &     &     &     & 3.2 &     &     &     & 3.8 &     &     &     &     &     &     &     &     &     &     &     &     & 14 \\ 
10 & COMBO-17           & 0.3 (0.02\%)& 1  &     &     &     & 0.3 &     &     &     &     &     &     &     &     &     &     &     &     &     &     &     &     &     &     &     \\ 
11 & DECaLS             & 670 (53\%)& 8  &     &     & 8.5 &     & 4.9 &     &     &     &     & 59 & 62 & 61 &     & 296 &     & 158 &     &     &     &     &     &     & 21 \\ 
12 & Legacy Survey      & 90 (7.1\%)& 9  &     &     & 11 &     &     & 3.3 & 13 & 8.9 &     &     &     &     &     &     & 8.9 & 38 & 0.2 &     &     &     & 5.8 & 0.7 &     \\ 
13 & DEEP2              & 2.5 (0.2\%)& 2  &     &     &     &     &     & 1.4 &     &     &     &     &     &     &     & 1.1 &     &     &     &     &     &     &     &     &     \\ 
14 & DES                & 595 (47\%)& 7  &     & 8.7 &     & 13 &     &     &     &     & 9.0 &     &     &     &     & 281 &     &     &     & 151 &     & 111 &     &     & 22 \\ 
15 & DataFusion-Spitzer & 66 (5.2\%)& 8  &     &     & 9.8 & 8.3 &     &     & 9.9 & 4.5 & 7.3 &     &     &     &     &     & 12 &     &     &     &     &     & 4.2 &     & 10 \\ 
16 & ESIS-VOICE         & 5.1 (0.4\%)& 1  &     &     &     &     &     &     &     &     & 5.1 &     &     &     &     &     &     &     &     &     &     &     &     &     &     \\ 
17 & FIREWORKS          & 0.1 (0.01\%)& 1  &     &     &     & 0.1 &     &     &     &     &     &     &     &     &     &     &     &     &     &     &     &     &     &     &     \\ 
18 & GOODS-ACS          & 0.1 (0.01\%)& 2  &     &     &     & 0.1 &     &     &     &     &     &     &     &     & 0.1 &     &     &     &     &     &     &     &     &     &     \\ 
19 & HSC                & 85 (6.7\%)& 8  &     &     &     &     & 5.1 & 1.3 & 7.7 &     &     & 19 & 13 & 17 &     & 7.8 &     &     &     &     &     &     &     &     & 14 \\ 
20 & Hawaii-HDFN        & 0.2 (0.02\%)& 1  &     &     &     &     &     &     &     &     &     &     &     &     & 0.2 &     &     &     &     &     &     &     &     &     &     \\ 
21 & IBIS               & 9.4 (0.7\%)& 1  &     &     & 9.4 &     &     &     &     &     &     &     &     &     &     &     &     &     &     &     &     &     &     &     &     \\ 
22 & INTWFC             & 42 (3.3\%)& 4  &     &     &     &     &     &     & 13 & 7.8 &     &     &     &     &     &     & 16 &     &     &     &     &     & 5.7 &     &     \\ 
23 & IRAC-EGS           & 0.5 (0.04\%)& 1  &     &     &     &     &     & 0.5 &     &     &     &     &     &     &     &     &     &     &     &     &     &     &     &     &     \\ 
24 & KIDS               & 269 (21\%)& 6  &     &     &     & 0.3 & 1.1 &     &     &     &     & 58 & 61 & 61 &     &     &     &     &     & 88 &     &     &     &     &     \\ 
25 & KPNO-FLS           & 9.1 (0.7\%)& 2  &     &     &     &     &     &     & 2.3 &     &     &     &     &     &     &     &     &     &     &     &     &     & 6.8 &     &     \\ 
26 & NDWFS              & 9.1 (0.7\%)& 1  &     &     & 9.1 &     &     &     &     &     &     &     &     &     &     &     &     &     &     &     &     &     &     &     &     \\ 
27 & PanSTARRS1-3SS     & 921 (72\%)& 18  & 9.2 &     & 11 & 13 & 5.1 & 3.4 & 13 & 9.1 &     & 62 & 63 & 62 & 0.7 & 362 & 21 & 178 &     & 79 & 0.1 &     & 7.4 &     & 22 \\ 
28 & RCSLenS            & 177 (14\%)& 4  &     &     &     &     &     &     & 14 & 7.8 &     &     &     &     &     & 134 & 21 &     &     &     &     &     &     &     &     \\ 
29 & SDSS-DR13          & 646 (51\%)& 15  & 2.6 &     & 11 &     & 5.1 & 3.5 & 13 & 9.1 &     & 62 & 63 & 62 & 0.7 & 363 & 21 &     &     &     & 0.1 &     & 7.4 &     & 22 \\ 
30 & SDSS-S82           & 115 (9.0\%)& 1  &     &     &     &     &     &     &     &     &     &     &     &     &     & 115 &     &     &     &     &     &     &     &     &     \\ 
31 & IAC-S82            & 113 (8.9\%)& 1  &     &     &     &     &     &     &     &     &     &     &     &     &     & 113 &     &     &     &     &     &     &     &     &     \\ 
32 & SDWFS              & 9.8 (0.8\%)& 1  &     &     & 9.8 &     &     &     &     &     &     &     &     &     &     &     &     &     &     &     &     &     &     &     &     \\ 
33 & SHELA              & 35 (2.8\%)& 1  &     &     &     &     &     &     &     &     &     &     &     &     &     & 35 &     &     &     &     &     &     &     &     &     \\ 
34 & SIMES              & 6.9 (0.5\%)& 1  &     & 6.9 &     &     &     &     &     &     &     &     &     &     &     &     &     &     &     &     &     &     &     &     &     \\ 
35 & SPLASH-SXDF        & 4.4 (0.3\%)& 1  &     &     &     &     &     &     &     &     &     &     &     &     &     &     &     &     &     &     &     &     &     &     & 4.4 \\ 
36 & SSDF               & 96 (7.6\%)& 1  &     &     &     &     &     &     &     &     &     &     &     &     &     &     &     &     &     &     &     & 96 &     &     &     \\ 
37 & SXDS               & 1.3 (0.1\%)& 1  &     &     &     &     &     &     &     &     &     &     &     &     &     &     &     &     &     &     &     &     &     &     & 1.3 \\ 
38 & SpARCS             & 33 (2.6\%)& 4  &     &     &     &     &     &     & 11 & 5.0 &     &     &     &     &     &     & 14 &     &     &     &     &     &     &     & 3.2 \\ 
39 & SpIES              & 74 (5.9\%)& 2  &     &     &     &     &     &     &     &     &     &     &     &     &     & 74 &     &     &     &     &     &     &     &     & 0.1 \\ 
40 & SpUDS              & 0.8 (0.1\%)& 1  &     &     &     &     &     &     &     &     &     &     &     &     &     &     &     &     &     &     &     &     &     &     & 0.8 \\ 
41 & UHS                & 65 (5.1\%)& 9  &     &     & 11 &     &     & 3.5 & 13 & 9.1 &     &     &     &     &     &     & 20 & 2.0 & 0.3 &     &     &     & 5.4 & 0.8 &     \\ 
42 & UKIDSS-DXS         & 23 (1.8\%)& 3  &     &     &     &     &     &     & 8.9 &     &     &     &     &     &     &     & 8.4 &     &     &     &     &     &     &     & 5.7 \\ 
43 & UKIDSS-LAS         & 602 (47\%)& 6  &     &     &     &     & 5.1 &     &     &     &     & 59 & 62 & 61 &     & 237 &     & 178 &     &     &     &     &     &     &     \\ 
44 & UKIDSS-UDS         & 0.9 (0.1\%)& 1  &     &     &     &     &     &     &     &     &     &     &     &     &     &     &     &     &     &     &     &     &     &     & 0.9 \\ 
45 & Ultradeep          & 0.4 (0.03\%)& 1  &     &     &     &     &     &     &     &     &     &     &     &     & 0.4 &     &     &     &     &     &     &     &     &     &     \\ 
46 & VICS82             & 81 (6.4\%)& 1  &     &     &     &     &     &     &     &     &     &     &     &     &     & 81 &     &     &     &     &     &     &     &     &     \\ 
47 & VIPERS-MLS         & 15 (1.1\%)& 1  &     &     &     &     &     &     &     &     &     &     &     &     &     &     &     &     &     &     &     &     &     &     & 15 \\ 
48 & VISTA-VHS          & 429 (34\%)& 7  &     & 8.7 &     & 8.7 &     &     &     &     & 9.0 & 24 &     &     &     & 247 &     &     &     &     &     & 110 &     &     & 22 \\ 
49 & VISTA-VIDEO        & 14 (1.1\%)& 3  &     &     &     & 5.1 &     &     &     &     & 3.6 &     &     &     &     &     &     &     &     &     &     &     &     &     & 5.2 \\ 
50 & VISTA-VIKING       & 436 (34\%)& 6  &     &     &     & 0.2 &     &     &     &     &     & 58 & 62 & 61 &     &     &     &     &     & 245 &     &     &     &     & 9.7 \\ 
51 & zBootes            & 6.7 (0.5\%)& 1  &     &     & 6.7 &     &     &     &     &     &     &     &     &     &     &     &     &     &     &     &     &     &     &     &     \\ 
\hline
\multicolumn{4}{l}{No surveys }    &  4   & 4 &  11   &  14   & 10    & 14    &  12   &  9   &  6   &  11   &  8   &   8  &  8   &  15   &   10  &    6  &   3  & 6 & 3 &   4  &  8   &   3  &  23   \\ 
\hline

\end{tabular}
\end{table}
\end{landscape}

\begin{landscape}
\thispagestyle{empty}
\begin{table}

\tiny
\centering
\caption{Overview of depths available in a given band. For each band we provide the area, $a$ in square degrees and the median value of $sigma$ averaged over the objetcs in every order 10 HEALPix cell in the field. These are 1 $\sigma$ depth values in $\mu$Jy in 2 arcsec apertures. Some K or Ks bands are empty because those fields are based on catalogues which only provide total fluxes. The small field SA13 (0.27 square degrees) is the only field with no K band coverage.}
\label{table:band_coverages}
\begin{tabular}{llllllllllllllllllllllllllllllllllllllllll }
\hline
field & \multicolumn{2}{c}{u}& \multicolumn{2}{c}{g}& \multicolumn{2}{c}{r}& \multicolumn{2}{c}{i}& \multicolumn{2}{c}{z}& \multicolumn{2}{c}{y}& \multicolumn{2}{c}{J}& \multicolumn{2}{c}{H}& \multicolumn{2}{c}{K}& \multicolumn{2}{c}{Ks}& \multicolumn{2}{c}{i1}& \multicolumn{2}{c}{i2}& \multicolumn{2}{c}{i3}& \multicolumn{2}{c}{i4} \\ 
\cline{2-29} 
& a & $\sigma$ & a & $\sigma$ & a & $\sigma$ & a & $\sigma$ & a & $\sigma$ & a & $\sigma$ & a & $\sigma$ & a & $\sigma$ & a & $\sigma$ & a & $\sigma$ & a & $\sigma$ & a & $\sigma$ & a & $\sigma$ & a & $\sigma$ \\ 
\hline
AKARI-NEP&&&9.6&1.5&9.6&1.6&9.6&1.5&9.6&3.2&9.6&7.2&&&&&&&&&7.2&1.7&7.2&1.4&&&&\\ 
AKARI-SEP&&&9.3&0.1&9.3&0.1&9.3&0.2&9.2&0.4&9.2&1.4&9.1&3.7&8.6&5.9&&&7.9&6.5&7.3&1.0&7.2&1.0&&&&\\ 
Bootes&&&11.9&0.1&11.9&0.1&11.9&0.2&11.9&0.2&11.9&4.3&11.8&1.4&9.7&1.5&9.7&3.0&5.3&12.1&10.1&0.8&10.1&1.0&10.0&5.4&10.0&7.4\\ 
CDFS-SWIRE&13.2&2.2&13.5&0.1&13.5&0.1&13.5&0.1&13.5&0.2&13.5&0.2&10.6&2.3&10.2&0.9&&&10.2&0.9&8.2&0.5&8.2&0.6&8.1&4.7&8.0&4.8\\ 
COSMOS&2.2&0.0&5.4&0.0&5.4&0.0&5.4&0.0&5.4&0.0&5.4&0.2&5.3&3.8&5.3&4.7&5.3&5.9&2.2&0.2&&&&&&&&\\ 
EGS&3.8&0.1&3.9&0.0&3.9&0.1&3.9&0.1&3.9&0.2&3.7&5.1&3.8&6.3&0.6&0.1&0.9&1.5&0.6&0.1&&&&&&&&\\ 
ELAIS-N1&13.2&0.0&13.9&0.0&13.9&0.0&13.9&0.1&13.9&0.0&13.8&0.2&9.3&0.5&&&9.3&0.7&&&9.7&0.8&9.7&1.0&9.6&4.9&9.7&4.9\\ 
ELAIS-N2&8.3&0.0&9.5&0.0&9.5&0.0&9.5&1.3&9.5&0.1&9.5&5.1&&&&&&&&&4.5&0.8&4.5&1.1&4.4&4.8&4.4&4.6\\ 
ELAIS-S1&&&9.5&0.1&9.4&0.1&9.5&0.2&9.5&0.4&9.5&1.1&9.4&2.6&8.2&3.7&&&9.2&5.3&7.2&0.7&7.4&1.0&6.7&5.3&6.7&5.5\\ 
GAMA-09&59.1&0.3&63.5&0.1&63.5&0.1&63.5&0.2&63.5&0.5&63.5&1.4&63.5&1.9&63.4&3.1&60.0&6.1&56.6&3.1&&&&&&&&\\ 
GAMA-12&62.1&0.2&64.2&0.1&64.2&0.1&64.2&0.4&64.2&0.5&64.1&1.4&64.1&1.7&64.0&2.7&62.4&6.3&61.8&2.9&&&&&&&&\\ 
GAMA-15&62.0&0.3&63.2&0.1&63.3&0.1&63.2&0.3&63.2&0.6&63.2&1.4&63.0&1.5&63.1&2.5&62.1&6.6&60.6&2.8&&&&&&&&\\ 
HDF-N&0.3&0.0&0.8&1.2&0.8&1.2&0.8&1.5&0.8&2.8&0.8&5.0&&&&&&&&&&&&&&&&\\ 
Herschel-Stripe-82&115.8&0.3&366.9&0.1&366.9&0.2&366.9&0.3&367.0&0.4&368.0&1.7&356.1&4.0&351.3&5.5&233.9&6.3&241.2&6.5&98.9&1.9&99.0&1.8&&&&\\ 
Lockman-SWIRE&14.8&0.0&22.1&0.0&22.1&0.0&22.1&1.2&22.1&0.0&22.0&6.2&20.4&5.5&&&8.6&0.6&&&11.6&0.7&11.6&0.9&11.5&5.0&11.5&5.3\\ 
NGP&&&179.6&1.1&179.6&1.1&179.6&1.2&179.6&1.1&179.6&3.3&177.2&4.1&177.3&5.1&178.4&5.7&&&&&&&&&&\\ 
SA13&&&0.3&0.2&0.3&0.3&&&0.3&1.0&&&0.3&6.8&&&&&&&&&&&&&&\\ 
SGP&291.8&1.8&298.7&0.1&290.1&0.1&292.8&0.3&294.7&0.5&272.4&1.3&244.2&1.5&226.6&2.4&&&226.2&2.6&&&&&&&&\\ 
SPIRE-NEP&&&0.2&1.6&0.2&1.6&0.2&1.6&0.2&3.9&0.2&7.1&&&&&&&&&&&&&&&&\\ 
SSDF&&&112.8&0.1&112.8&0.1&112.8&0.2&112.8&0.4&112.7&1.2&111.7&2.7&98.0&4.1&&&107.5&5.5&94.3&79.7&93.0&86.9&&&&\\ 
xFLS&4.7&2.4&7.8&0.2&7.9&0.1&7.8&1.5&7.8&2.5&7.8&6.2&5.6&6.1&&&&&&&4.1&5.4&4.1&6.0&4.0&21.0&4.0&20.4\\ 
XMM-13hr&&&0.8&0.2&0.8&0.2&&&0.8&0.8&&&0.9&6.2&&&&&&&&&&&&&&\\ 
XMM-LSS&18.9&0.1&22.4&0.0&22.5&0.0&22.5&0.0&22.4&0.1&22.4&0.2&22.3&2.6&21.1&3.6&7.1&0.6&22.0&5.1&10.0&0.7&9.8&0.8&9.3&5.8&9.3&6.4\\ 
 \hline

\end{tabular}
\end{table}
\end{landscape}

\section{Data access}
\label{sec:data_access}

All the notebooks used to perform the reduction are available here: \\

\url{https://github.com/H-E-L-P/dmu_products/}\\

These notebooks make use of the following \texttt{Python} package:\\

\url{https://github.com/H-E-L-P/herschelhelp_internal}\\

Some code to aid accessing and using the database is available here:\\

\url{https://github.com/H-E-L-P/herschelhelp_python}\\

The database is structured in line with the code on GitHub in order to make relative links persistent. The folder dmu\_products/dmu0/ contains the pristine catalogues and dmu\_products/dmu1/ contains the standardised and merged catalogues. The data is available to download here:\\

\url{http://hedam.lam.fr/HELP/dataproducts/}\\

The data is also available via Virtual Observatory \citep[VO;][]{Demleitner:2014} protocols at the dedicated server the Virtual Observatory at susseX (VOX):\\

\url{https://herschel-vos.phys.sussex.ac.uk/}\\

This permits quick searching and downloading of small subsets including programmatic access. A full description of the over one thousand column headings can be found here:\\

\url{https://herschel-vos.phys.sussex.ac.uk/herschelhelp/q/cone/info}\\

Examples of accessing the database with Python, including all the code used to produce the figures presented in this paper, can be found here:\\

\url{https://github.com/H-E-L-P/dmu_products/tree/master/dmu31/dmu31_Examples}\\

There is also an all sky viewer for viewing the catalogue over Herschel imaging or other imaging surveys:\\

\url{http://hedam.lam.fr/HELP/dataproducts/dmu31/dmu31_HiPS/viewer/}\\

\section{Diagnostic plots}
\label{sec:diagnostics}
\begin{figure}
\centering
\includegraphics[width=0.4\textwidth]{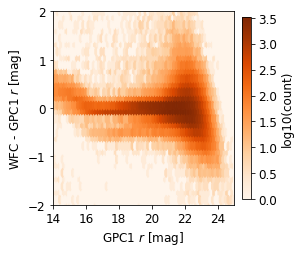}
\caption{\label{fig:mag_compare} Comparison between PanSTARRS GPC1 r band magnitudes and INT-WFC r band magnitudes on the example field ELAIS-N1. We see errors increasing for fainter objects.}
\end{figure}

\begin{figure}
\centering
\includegraphics[width=0.35\textwidth]{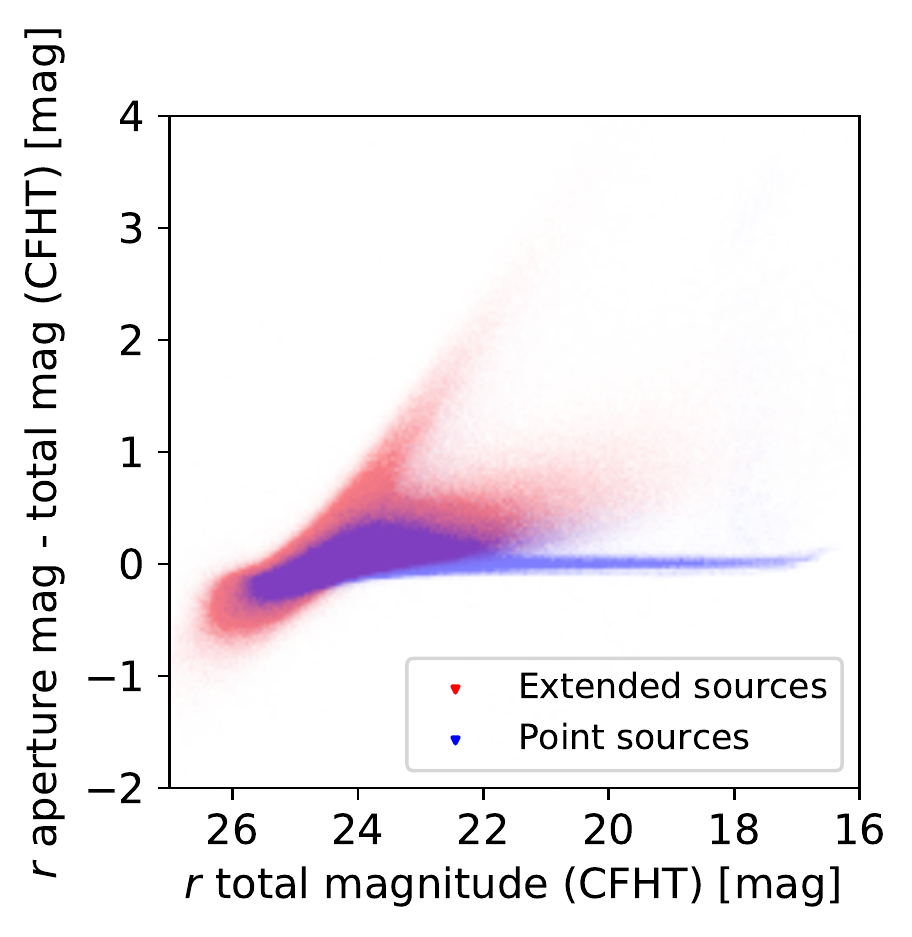}
\caption{\label{fig:ap_vs_tot} Comparison between aperture and total magnitude for CFHT r band on the example field ELAIS-N1. It is shown separately for extended and point sources. Unsurprisingly, point sources tend to equality and extended objects show differences. The small number of spurious objects at the bright end are due to saturation in the CFHT imaging. A feature which is also captured in figures \ref{fig:outliers}, \ref{fig:flags}, and \ref{fig:numbers_g_allfields}.}
\end{figure}

We here provide figures~\ref{fig:mag_compare} and \ref{fig:ap_vs_tot} which are described in section~\ref{sec:pipeline} and used for validating the final catalogue. These are automatically generated for every field and many combinations of bands and can be seen in the notebooks on GitHub. The diagnostics for the example field ELAIS-N1, for instance, are viewable here:\\

\url{http://hedam.lam.fr/HELP/dataproducts/dmu1/dmu1_ml_ELAIS-N1/notebook_output/3_Checks_and_diagnostics.html}\\

\bsp	

\end{document}